\documentclass[11pt]{article}
\def\dofigures{1}
\pdfoutput=1

\usepackage{epsfig,multicol,bbm,simplewick}

\usepackage[usenames]{color}
\usepackage{fix-cm}
\usepackage{tensor}
\usepackage{booktabs}
\usepackage[pdftex,
  pdftitle={A Class of Effective Field Theory Models of Cosmic Acceleration},
  pdfauthor={Jolyon K Bloomfield, Eanna E Flanagan},
  bookmarks,bookmarksopen=false,
  pdfstartview={FitH}
]{hyperref}

\newcommand{\Feyn}[1]{#1\kern-0.45em/}

\usepackage[english]{babel}
\usepackage{amsmath,amssymb,amsbsy,amstext}
\usepackage{graphicx}
\usepackage{amsfonts}
\usepackage{amssymb}
\usepackage{subfigure}
\usepackage{color}
\usepackage{float}
\usepackage[small]{caption}

\numberwithin{equation}{section}

\def\bes{\begin{subequations}}
\def\ees{\end{subequations}}

\def\mpl{m_{p}}

\DeclareRobustCommand{\SkipTocEntry}[4]{}

\begin{document}

\begin{titlepage}

% ----------------------------------------------------------------------
%
% TIME OF DAY
%
\newcount\hh
\newcount\mm
\mm=\time
\hh=\time
\divide\hh by 60
\divide\mm by 60
\multiply\mm by 60
\mm=-\mm
\advance\mm by \time
\def\hhmm{\number\hh:\ifnum\mm<10{}0\fi\number\mm}

% ----------------------------------------------------------------------

\setcounter{page}{1} \baselineskip=15.5pt \thispagestyle{empty}

\bigskip\
\begin{center}
{\fontsize{17.5}{30}\selectfont  \bf A Class of Effective Field Theory Models of Cosmic Acceleration}
\end{center}

\vspace{0.5cm}
\begin{center}
{\fontsize{14}{30}\selectfont  Jolyon K. Bloomfield,$^{1,2,3}$ \'Eanna
  \'E. Flanagan,$^{1,2,4}$}
%{\fontsize{14}{30}\selectfont  Jolyon K. Bloomfield,$^{1,2}$ \'Eanna \'E. Flanagan,$^{1,2}$}
\end{center}

%\vspace{0.2cm}

\begin{center}
\vskip 8pt
\textsl{${}^1$ Center for Radiophysics and Space Research, Cornell University, Ithaca, NY 14853.}

\vskip 4pt
\textsl{${}^2$ Laboratory for Elementary Particle Physics, Cornell University, Ithaca, NY 14853.}

\vskip 4pt
\textsl{${}^3$ jkb84@cornell.edu}

\vskip 4pt
\textsl{${}^4$ eef3@cornell.edu}

\end{center} %\vfil

%\vspace{0.8cm}

\vskip 1cm
\begin{center}
%{\it Draft of 25 January 2012; printed \today{} at \hhmm}
{\it $25^{th}$ January 2011}
\end{center}

\vspace{1.2cm}
\hrule \vspace{0.3cm}
{ \noindent \textbf{Abstract} \\[0.2cm]
\noindent
We explore a class of effective field theory models of cosmic acceleration
involving a metric and a single scalar field.  These models can be obtained by starting with a set of ultralight
pseudo-Nambu-Goldstone bosons whose couplings to matter satisfy the
weak equivalence principle, assuming that one boson is lighter than
all the others, and integrating out the heavier fields.
The result is a quintessence model with matter coupling,
together with a series of correction terms in the action in a covariant derivative expansion, with specific scalings for the coefficients.
After eliminating higher derivative terms and exploiting the field redefinition freedom,
we show that the resulting theory contains nine independent free functions of the scalar field when truncated at four derivatives.  This is in contrast to the four free functions found in
similar theories of single-field inflation, where matter is not
present.  We discuss several different representations of the theory
that can be obtained using the field redefinition freedom.
For perturbations to the quintessence field today on subhorizon
lengthscales larger than the Compton wavelength of the heavy fields,
the theory is weakly coupled and natural in the sense of t'Hooft.  The
theory admits a regime where the perturbations become modestly nonlinear, but very strong
nonlinearities lie outside its domain of validity.}
 \vspace{0.3cm}
 \hrule

\vspace{0.6cm}

\end{titlepage}

\newpage
\tableofcontents

\eject
\section{Introduction and Summary}
\label{sec:intro}

\subsection{Background and Motivation}

The recent discovery of the accelerating expansion of the
Universe \cite{Perlmutter1999, Riess1998} has prompted
many theoretical speculations about the underlying mechanism.
The most likely mechanism is a cosmological constant, which is the
simplest model and is in good agreement with observational data
\cite{Blake2011}.  More complicated models involve new dynamical
sources of gravity that act as dark energy, and/or modifications to
general relativity on large scales.
A plethora of
models have been postulated and explored in recent years, including
Quintessence, K-essence
\cite{Armendariz-Picon2000, Chiba2000}, Ghost Condensates
\cite{Arkani-Hamed2004}, DGP gravity \cite{Dvali2000}, and $f(R)$ gravity, to
name but a few.  See Refs.\ \cite{Peebles2003, Nobbenhuis2006, Copeland2006,
Caldwell:2009ix,Silvestri:2009hh,Skordis2011,Amendola2010} for detailed reviews of
these and other models.

A common feature of the majority of dark energy and modified gravity models is that in
the low energy limit, they are equivalent to
general relativity coupled to one or more scalar fields, often called
quintessence fields.
Therefore it is useful to try to construct very general low energy
effective quantum field theories of general relativity coupled to
light scalar fields, in order to encompass broad classes of dark
energy models.
Considering dark energy models as quantum field theories is useful,
even though the dynamics of dark energy is likely in a classical
regime, because it facilitates discriminating against theories which
are theoretically inconsistent or require fine tuning.

A similar situation occurs in the study of models of inflation,
where it is useful to construct generic theories using effective field
theory.
Cheung \textit{et al.} \cite{Cheung2008} constructed a
general effective field theory for gravity and a single inflaton
field, for perturbations about a background Friedman-Robertson-Walker
cosmology in unitary gauge.  This work was later generalized in
multiple directions
\cite{Senatore:2010wk,2011arXiv1106.2189F} and has been very useful.
An alternative approach to single field inflationary models was taken
by Weinberg \cite{Weinberg2008}, who constructed an effective field
theory to describe both the background cosmology and the
perturbations.  This theory consisted at leading order of a standard
single field inflationary model with a potential, together with
higher order terms in a covariant derivative expansion up to
four derivatives.  More detailed discussions of this type of effective
field theory were given by Burgess, Lee and Trott \cite{Burgess:2009ea}.

When one turns from inflationary effective field theories to
quintessence effective field theories, the essential physics is very
similar, but there are three important differences that arise:

\begin{itemize}

\item First, the hierarchy of scales is vastly more extreme in
  quintessence models.  The Hubble parameter $H$ is typically several
  orders of magnitude below the Planck scale $\mpl \sim 10^{28}$ eV in inflationary
  models, whereas for quintessence models $H_0 \sim 10^{-33} \, $eV is
  $\sim 60$ orders of magnitude below the Planck scale.  Quintessence
  fields must have a mass that is smaller than or on the order of
  $H_0$.  It is a well-known, generic challenge for quintessence
  models to ensure that loop effects do not give rise to a mass much
  larger than $H_0$.  Because of the disparity of scales, this issue
  is more extreme for quintessence models than inflationary models.

\item In most inflationary models, it is assumed that the dynamics of
  the Universe are dominated by gravity and the scalar field (at least
  until reheating).  By contrast, for quintessence models in the
  regime of low redshifts relevant to observations, we know
  that cold dark matter gives an $O(1)$ contribution to the energy
  density.  Therefore there are additional possible couplings and
  terms that must be included in an effective field theory.

\item For any effective field theory, it is possible to pass outside
the domain of validity of the theory even at energies $E$ low compared to
the theory's cutoff $\Lambda$, if the mode occupation numbers $N$ are
sufficiently large (see Sec.\ \ref{sec:validity} below for
more details).  This corresponds to a breakdown of the classical
derivative expansion.
For quintessence theories, mode occupation numbers today
can be as large as $N \sim (\mpl/H_0)^2$ and it is possible to pass
outside the domain of validity of the theory.  By contrast in
inflationary models, this is less likely to occur since mode
occupation numbers for the perturbations are not large before modes exit the horizon.
Thus, the effective field theory framework is less all-encompassing
for quintessence models than for inflation models.
This issue seems not to have been appreciated in the literature and we discuss it
in Sec. \ref{sec:validity} below.

\end{itemize}

Several studies have been made of generic effective field theories of
dark energy.  Creminelli, D'Amico, Nore\~{n}a and Vernizzi
\cite{Creminelli2009} constructed a the general effective theory of
single-field quintessence for perturbations about an arbitrary FRW
background, paralleling the similar construction for inflation
\cite{Cheung2008}.  Park, Watson and Zurek constructed an effective
theory for describing both the background cosmology and the
perturbations, following the approach of Weinberg \cite{Weinberg2008}
but generalizing it to include couplings to matter \cite{Watson2010}.

The two approaches to effective field theories of quintessence --
specialization to perturbations about a specific background,
 and maintaining covariance and the ability to describe the dynamics
 of a variety of backgrounds -- are complementary to one another.
The dynamics of the cosmological background FRW solution
can be addressed in the covariant approach of Weinberg,
but not in the background specific approach of Creminelli et al.,
which restricts attention to the dynamics of perturbations about a
given, fixed background. On the other hand, a background specific approach can describe a larger set of dynamical theories for the perturbations than can a covariant derivative expansion\footnote{To see this, consider for example a term in the
  Lagrangian of the form $f(\phi) (\nabla \phi)^{2 n}$, where $\phi$
  is the quintessence field.  Such a term
  would be omitted in the covariant derivative expansion for
  sufficiently large $n$.  However, upon expanding this term using $\phi = \phi_0 +
  \delta \phi$, where $\phi_0$ is the background solution, one finds
  terms $\sim (\nabla \phi_0)^{2n-2} (\nabla \delta \phi)^2$ which are
  included in the Creminelli approach of applying standard effective
  field theory methods to the perturbations.}.

\subsection{Approach and Assumptions}
\label{sec:approach}

The purpose of this paper is to revisit, generalize and correct
slightly the covariant effective field theory analysis of
Park, Watson and Zurek \cite{Watson2010}.
Following Weinberg and Park et al., we restrict attention to theories
where the only dynamical degrees of freedom are a graviton and a
single scalar.  We allow couplings to an arbitrary matter sector, but
we assume the validity of the weak equivalence principle, motivated by
the strong experimental evidence for this principle.
We assume that the theory consists of a standard quintessence theory coupled to matter at leading
order in a derivative expansion, with an action of the form
\begin{align}
  S[g_{\alpha\beta},\phi,\psi_{\rm m}] = \int d^4 x \sqrt{-g} \left\{
    \frac{\mpl^2}{2} R - \frac{1}{2} (\nabla \phi)^2 - U(\phi)
  \right\} + S_{\rm m} \left[ e^{\alpha(\phi)} g_{\mu \nu},  \psi_{\rm
    m}\right]. \label{eq:intro:leadingorder}
\end{align}
Here $\psi_{\rm m}$ denotes a set of matter fields, and $\mpl$ is the
Planck mass. The factor $e^{\alpha(\phi)}$ in the matter action
provides a leading-order non-minimal coupling of the quintessence
field to matter, in a manner similar to Brans-Dicke models in the
Einstein frame.

Our analysis then consists of a series of steps:

\begin{enumerate}

\item We add to the action all possible terms involving the scalar
  field and metric, in a covariant derivative expansion up to four
  derivatives. We truncate the expansion at four derivatives, as this
  is sufficient to yield the leading corrections to the
  action (\ref{eq:intro:leadingorder}).  As described by Weinberg
  \cite{Weinberg2008} there are
  ten possible terms, with coefficients that can be arbitrary
  functions of $\phi$ [see Eq.\ (\ref{eq:S1}) below].
  Section \ref{sec:scaling} below describes one possible justification
of this covariant derivative expansion from an effective field theory
viewpoint, starting from a set of ultralight pseudo Nambu-Goldstone
bosons (PNGBs).  It is likely that the same expansion can be obtained
from other, more general starting points.

\item We allow for corrections to the coupling to matter by
adding to the metric that appears in the matter action
all possible terms involving the metric and $\phi$ allowed by the
derivative expansion, that is, up to two derivatives.  There are six
such terms [see Eq.\ (\ref{eq:JordanMetric}) below.]
We also add to the action terms involving the stress energy tensor
$T_{\mu\nu}$ of
the matter fields, up to the order allowed by the derivative expansion
using $T_{\mu\nu} \sim \mpl^2 G_{\mu\nu}$ [see Eq.\ (\ref{eq:S1}) below].
Including such terms in the action seems poorly motivated, since {\it a
priori} there is no reason to expect that the resulting theory would
respect the weak equivalence principle (see Appendix \ref{app:WEP}).
However we show in Appendix \ref{app:WEP} that the weak equivalence
principle is actually satisfied, to the order we are working to in the
derivative expansion.  In addition, all the terms in the action
involving $T_{\mu\nu}$ can be shown to have equivalent representations
not involving the stress energy tensor, using field redefinitions (see
Appendix \ref{app:WEP}).

\item The various correction terms are not all independent because of
  the freedom to perform field redefinitions involving $\phi$,
  $g_{\mu\nu}$ and the matter fields, again in a derivative expansion.  In Sec.\
  \ref{sec:techs} we explore the space of such field redefinitions,
finding eleven independent transformations and tabulating their effects on the
coefficients in the action (see Table \ref{tab:transformations} below).

\item Several of the correction terms that are obtained from the
  derivative expansion are ``higher derivative'' terms, by which we
  mean that they give contributions to the equations of motion which
  involve third-order or higher-order time derivatives of the
  fields\footnote{The precise definition of higher derivative that we
    use, which is covariant, is that an equation will be said not to
    contain any higher derivative terms if there exists a choice of
    foliation of spacetime for which any third-order or higher-order
    derivatives contain at most two time derivatives.
    Theories which are
    higher derivative in this sense are generically associated with
    instabilities (Ostragradski's theorem) \cite{Woodard2007},
    although the instabilities can be evaded in special cases, for
    example $f(R)$ gravity.  For most of this paper (except for the
    Chern-Simons term), a simpler definition of higher derivative would
    be sufficient: a term in the action is ``higher derivative'' if it
    gives rise to terms in the equation of motion that involve any
    third-order or higher order derivatives.}.
Normally, such higher derivative terms give rise to additional degrees of freedom.
However, if they are treated
perturbatively (consistent with our derivative expansion) additional
degrees of freedom do not arise.  Specifically,
one can perform a {\it reduction of order} procedure on the equations of
motion \cite{1971ctf..book.....L,Parker:1993dk,Flanagan:1996gw},
substituting the zeroth-order equations of motion
into the higher derivative terms in the equations of motion to
eliminate the higher derivatives\footnote{This is more general than
  requiring the solutions of the equation of motion to be analytic in
  the expansion parameter, as advocated by Simon \protect{\cite{Simon1990}}; see
  Ref.\ \protect{\cite{Flanagan:1996gw}}.}.
We actually use a slightly different but equivalent procedure of
eliminating the higher derivative terms directly in the action using
field redefinitions\footnote{This procedure is counterintuitive since
  normally field redefinitions
do not change the physical content of a theory; here however they do
because the field redefinitions themselves involve higher derivatives.}
 (see Appendix \ref{app:backsubs}).

Weinberg \cite{Weinberg2008} and Park {\it et al.} \cite{Watson2010}
use a slightly different method, consisting of substituting the leading order
equations of motion directly into the higher derivative terms in the
action.  This method is not generally valid, but it is valid up to field
redefinitions that do not involve higher derivatives, and so it
suffices for the purpose of attempting to classify general theories of
dark energy (see Appendix \ref{app:backsubs}).

\item Another issue that arises with respect to the higher derivative
  terms is the following.  Is it really necessary to include such
  terms in an action when trying to write down the most general theory
  of gravity and a scalar field, in a derivative expansion?
Weinberg \cite{Weinberg2008} suggested that  perhaps a more general class of
theories is generated by including these terms and performing a
reduction of order procedure on them, rather than by omitting them.
However, since it is ultimately possible to obtain a theory that is
perturbatively equivalent to the higher derivative theory, and which
has second order equations of motion, it should be possible just to
write down the action for this reduced theory.  In other words,
an equivalent class of theories should be obtained simply by omitting
all the higher derivative terms from the start.  We show explicitly
in Sec.\ \ref{sec:details} that this is the case for the class of
theories considered here.

\item We fix the remaining field redefinition freedom by choosing a
  ``gauge'' in field space, thus fixing the action uniquely (see
  Sec. \ref{sec:details:final}).

\end{enumerate}

\subsection{Results and Implications}
Our final action is [Eq.\ (\ref{eq:details:finalaction}) below]
\begin{align}
S =&  \int d^4 x \sqrt{-g}
\left\{ \frac{\mpl^2}{2} R - \frac{1}{2}
  (\nabla \phi)^2 - U(\phi)\right\}
+ S_{\rm m}[e^{\alpha(\phi)} g_{\alpha\beta},\psi_{\rm m}] \nonumber \\
&
+ \epsilon \int d^4 x \sqrt{-g}  \bigg\{a_1 (\nabla \phi)^4
+  b_2 T (\nabla \phi)^2
+  c_1 G^{\mu \nu} \nabla_\mu \phi \nabla_\nu \phi
\nonumber \\
& +  d_3 \left( R^2 - 4 R^{\mu \nu} R_{\mu \nu} + R_{\mu \nu \sigma \rho} R^{\mu \nu \sigma \rho} \right)
%\nonumber \\ &&
  +  d_4 \epsilon^{\mu \nu \lambda \rho} \tensor{C}{_{\mu
    \nu}^{\alpha \beta}} C_{\lambda \rho \alpha \beta} + e_1 T^{\mu
  \nu} T_{\mu \nu} + e_2 T^2 + \ldots
\bigg\}.
\label{eq:details:finalaction0}
\end{align}
Here the coefficients $a_1$, $b_2$ etc.\ of the
next-to-leading order terms in the derivative
expansion are arbitrary functions of $\phi$, and the ellipsis $\ldots$
refers to higher order terms with more than four derivatives.
The corresponding equations of motion do
not contain any higher derivative terms.  This result generalizes that
of Weinberg \cite{Weinberg2008} to include couplings to matter.

\begin{figure}[h!]
    \centering
\ifx\dofigures\undefined
\else
        \includegraphics[width=0.55\textwidth]{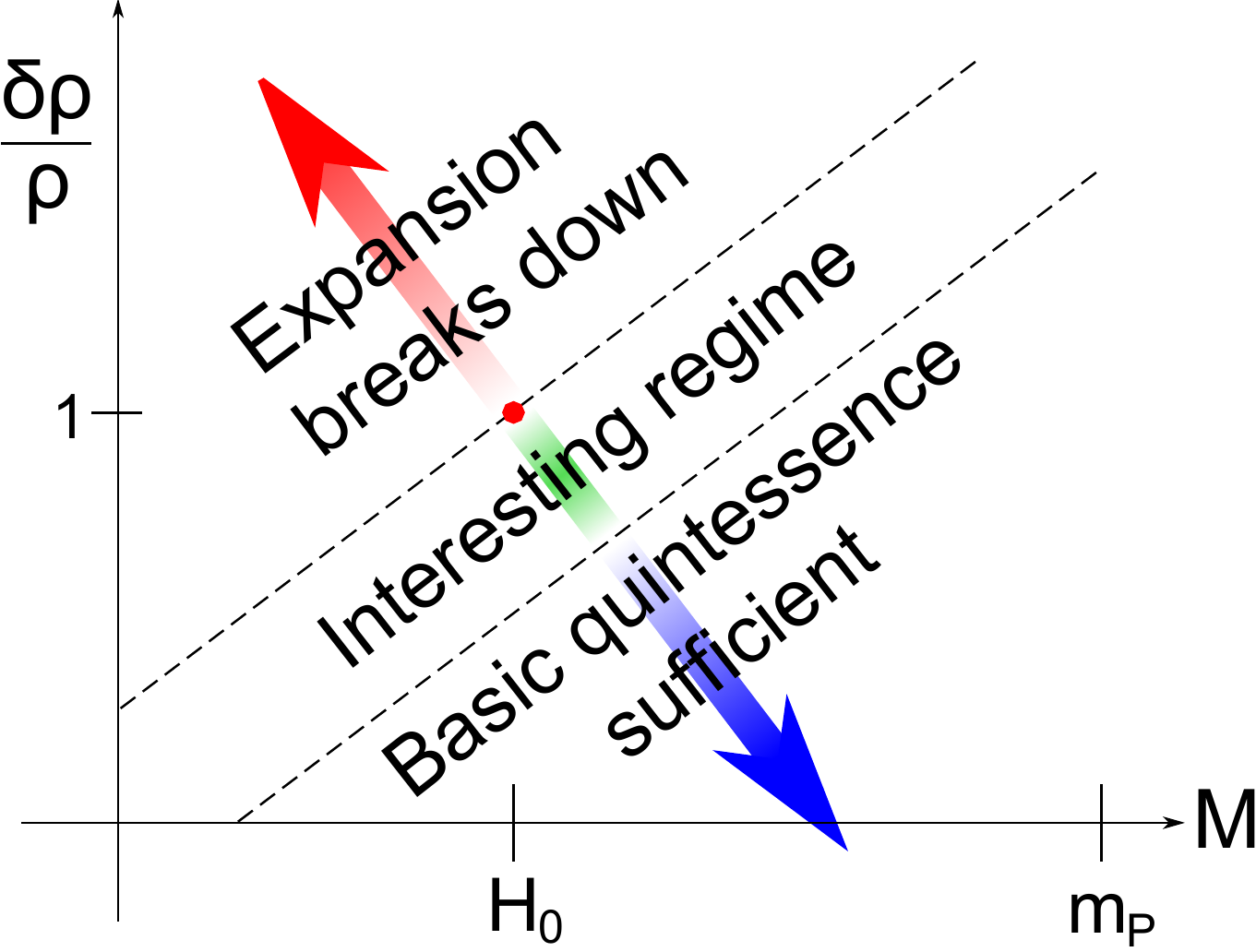}
\fi
    \caption{\sl The parameter space of fractional density
      perturbation $\delta \rho/\rho$ for perturbations to the
      quintessence field, and cutoff scale $M$ for the effective field
    theory, illustrating the constraint (\ref{eq:ddd}) on the domain
    of validity.  Near the boundary of the domain of validity the
    higher derivative terms in the action are potentially observable,
    this is labeled the ``interesting regime''.  Further away from the
    boundary the higher derivative terms are negligible and the theory
    reduces to a standard quintessence model with a matter coupling.}
    \label{fig:scales0}
\end{figure}

\begin{figure}[h!]
    \centering
\ifx\dofigures\undefined
\else
        \includegraphics[width=0.80\textwidth]{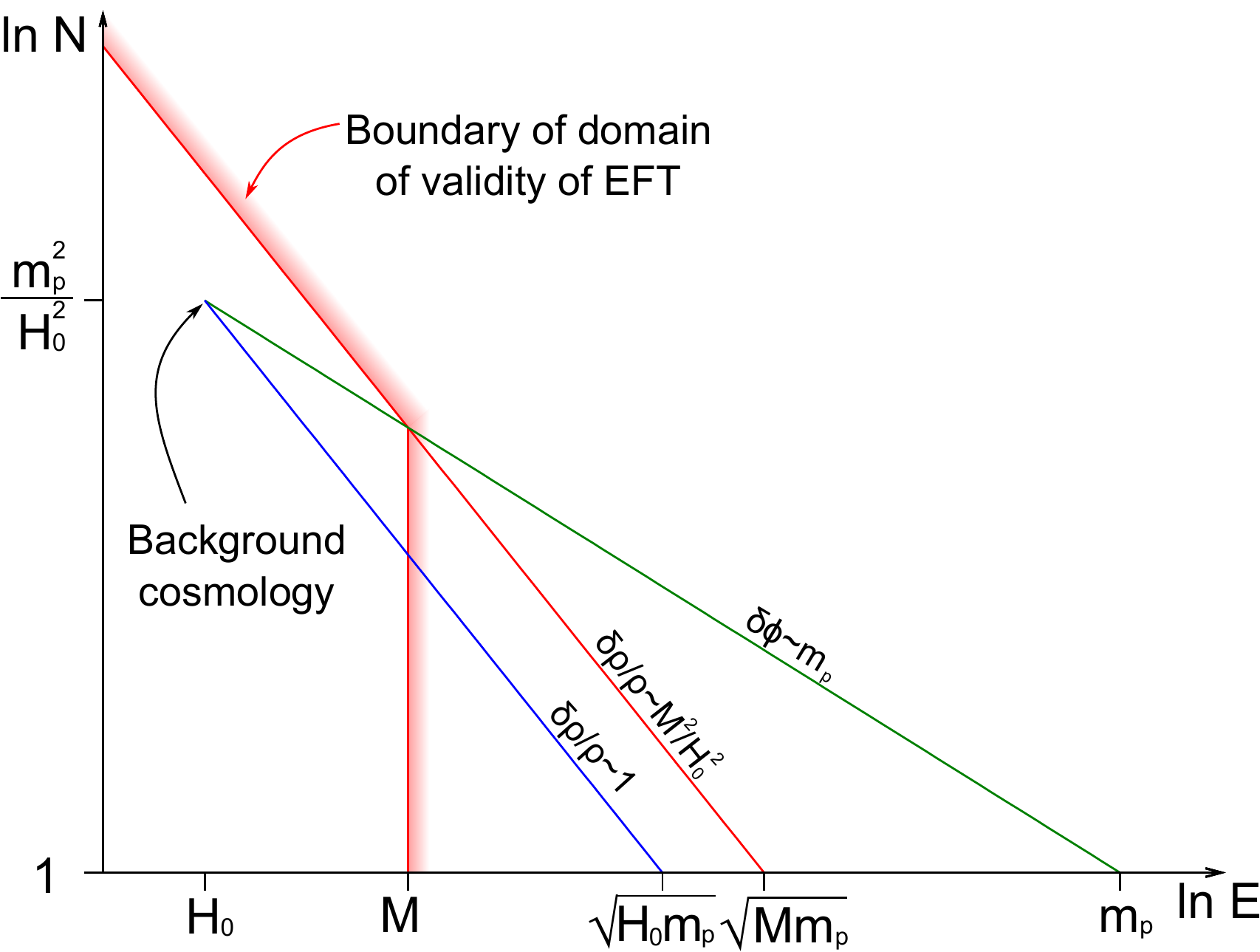}
\fi
    \caption{\sl The domain of validity of the effective field theory
      in the two dimensional parameter space of energy $E$ per quantum
      of a mode of the quintessence field, and mode occupation number
      $N$. The cutoff scale $M$ must be larger than the Hubble
      parameter $H_0$ in order that the background cosmology lie
      within the domain of validity. Perturbation modes on
      length-scales that are small compared to $H_0^{-1}$ but large
      compared to $M^{-1}$ can be described, but only if the mode
      occupation number and fractional density perturbation are
      sufficiently small.  See Sec.\ \ref{sec:validity} for details.}
    \label{fig:scales1}
\end{figure}

We can summarize our key results as follows:

\begin{itemize}

\item The most general action contains nine free functions of $\phi$:
  $U, \alpha, a_1, b_2, c_1, d_3, d_4, e_1, e_2$, as compared to the
  four functions that are needed when matter is not present
  \cite{Weinberg2008}.

\item There are a variety of different forms of the final theory that
  can be obtained using field redefinitions.  In particular some
of the matter-coupling terms in the action can be re-expressed as
terms that involve only the quintessence field and metric.
Specifically, the term $T (\nabla \phi)^2$ term could be eliminated in
favor of $\square \phi (\nabla \phi)^2$, the $(\nabla \phi)^4$ could
be eliminated in favor
of a term $T^{\mu\nu} \nabla_\mu \phi \nabla_\nu \phi$, or the
$G^{\mu\nu} \nabla_\mu \phi
  \nabla_\nu \phi$ term could be eliminated in favor of a term
$T^{\mu\nu} \nabla_\mu \phi \nabla_\nu \phi$ (see Sec.\
\ref{sec:details:final}).

\item As mentioned above, one obtains the correct final action if one
  excludes throughout the calculation all higher derivative terms.

\item The final theory does contain terms involving the matter
  stress-energy tensor.  Nevertheless, the weak equivalence principle
  is still satisfied (see Appendix \ref{app:WEP}).  It is possible
to eliminate the stress-energy terms, but only if we allow
higher derivative terms in the action (where it is assumed that the
reduction of order procedure will be applied to these higher
derivative terms).  Thus, for a fully general theory, one must have
either stress-energy terms or higher derivative terms; one cannot
eliminate both (see Sec.\ \ref{sec:details:final}).

\item We can estimate how all the coefficients $a_1$ etc. scale with
  respect to a cutoff scale $M$ for an effective field theory as
  follows (see Sec.\ \ref{sec:scaling}).  We assume that several
  ultralight scalar fields of mass $\sim H_0$ arise as pseudo
  Nambu-Goldstone bosons (PNGBs) from some high energy
  theory \cite{1995PhRvL..75.2077F,2003JCAP...07..003A}, and are
  described by a nonlinear sigma model at low energies.  We then
  suppose that all but one of the these PNGB fields have masses $M$ that are
  somewhat larger than $\sim H_0$, and integrate them out.  This will
  give rise to a theory of the form discussed above for the single light
  scalar, where the higher derivative terms are suppressed by powers
  of $M$.  The scalings for each of the coefficients in the action
  are summarized in Table \ref{tab:scalings}.  We find that the
  fractional corrections to the cosmological
  dynamics due to the higher derivative terms scale as $H_0^2/M^2$,
  as one would expect.

\item Finally, we can use these scalings to estimate the domain of
  validity of the effective field theory (see Sec.\ \ref{sec:validity}).  We find
  that cosmological
  perturbations with a density perturbation $\delta \rho$ in the
  quintessence field must have a fractional density perturbation that
  satisfies
\begin{align}
\frac{\delta \rho}{\rho} \ll \frac{M^2}{H_0^2}.
\label{eq:ddd}
\end{align}
Thus perturbations can become nonlinear, but only modestly so, if $M$
is close to $H_0$.  The parameter space of fractional density
perturbation $\delta \rho/\rho$ and cutoff scale $M$ is illustrated in
Fig.\ \ref{fig:scales0}.  In addition there is the standard constraint
for derivative expansions
\begin{align}
E \ll M
\label{eq:ddd1}
\end{align}
where $E^{-1}$ is the length-scale or time-scale for some process.
We show in Fig.\ \ref{fig:scales1} the two constraints (\ref{eq:ddd}) and
(\ref{eq:ddd1}) on the two
dimensional parameter space of energy $E$ and mode occupation number $N$.

\end{itemize}

Finally, in Appendix \ref{app:compare} we compare our analysis to
that of Park, Watson and Zurek
\cite{Watson2010}, who perform a similar computation but in the Jordan
frame rather than the Einstein frame (see also Ref.\
\cite{Jimenez:2011nn}).
The main difference between our
analysis and theirs is that they
use a different method to estimate the scalings of the coefficients,
and as a result their final action
differs from ours, being parameterized by three free functions rather
than nine.

%\eject

\ifx\doejects\undefined
\else
        \eject
\fi

\iffalse
\subsection{Outline of this Paper}

We now move on to the detailed calculation of our results. The
remainder of this paper is structured as follows. In section
\ref{sec:conformal}, we discuss the issue of the choice of conformal
frame and how it affects our analysis. Section \ref{sec:validity}
looks at the regime of validity of the theory, while Section
\ref{sec:techs} introduces the techniques we use to work with the
theory. We perform the bulk of our analysis in Section
\ref{sec:details}. We discuss other issues and conclude in Section
\ref{sec:conclusion}.

Our notations and conventions are given in Appendix
\ref{app:notation}. In Appendix \ref{app:transform}, we describe some
details for the transformations we use. A short analysis of the
implications of including terms involving the stress-energy tensor in
the action on the weak equivalence principle is presented in Appendix
\ref{app:WEP}. Finally, we present a derivation of the reduction
technique used by Weinberg and Park et al in terms of our techniques
in Appendix \ref{app:backsubs}.
\fi

\section{Class of Theories Involving Gravity and a Scalar Field}
\label{sec:definetheory}

As discussed in the Introduction, our starting point is an action for a standard quintessence model with an arbitrary matter
coupling, together with a perturbative correction which consists of a
general derivative expansion up to four derivatives.  The action
is a functional of the Einstein-frame metric $g_{\alpha\beta}$, the
quintessence field $\phi$, and some matter fields which we denote collectively by
$\psi_{\rm m}$:
\begin{align}
S[g_{\alpha\beta},\phi,\psi_{\rm m}] = S_0[g_{\alpha\beta},\phi] +
\epsilon S_1[g_{\alpha\beta},\phi,T_{\alpha\beta}(\psi_{\rm m})]
+ S_{\rm m}[{\bar g}_{\alpha\beta},\psi_{\rm m}] + O(\epsilon^2).
\label{eq:start}
\end{align}
Here $S_{\rm m}$ is the action for the matter fields, and the quantity
$\epsilon$ is a formal expansion parameter.
We will see in Sec.\ \ref{sec:scaling} below that $\epsilon$ can be identified
as proportional to $M^{-2}$, where $M$ is a cutoff scale or the mass of the lightest
of the fields that have been integrated out to obtain the low energy
action.  Equivalently, $\epsilon$ counts the number of derivatives in
our derivative expansion, with $\epsilon^n$ corresponding to $2n$ derivatives.
The notation
in the second term indicates that the perturbative correction $S_1$ to the
action can depend on the matter fields, but only through their stress
energy tensor $T_{\alpha\beta}$ (defined in Appendix \ref{app:notation}).
Explicitly we have
\begin{align}
S_0 = \int d^4 x \sqrt{-g}  \left[ \frac{\mpl^2}{2} R - \frac{1}{2}
  (\nabla \phi)^2 - U(\phi)\right],
\label{eq:S0}
\end{align}
and \cite{Weinberg2008,Watson2010}
%
%  29 Oct 2011
% OLD: b5 e3 e4
% NEW: b7 b5 b6
%
\begin{align}
S_1 = \int d^4 x \sqrt{-g} & \bigg\{a_1 (\nabla \phi)^4 +  a_2 \square
  \phi (\nabla \phi)^2 +  a_3 (\square \phi)^2
 \nonumber \\ &
+  b_1 T^{\mu \nu} \nabla_\mu \phi \nabla_\nu \phi +  b_2 T (\nabla \phi)^2 +  b_3 T \square \phi +
 b_4 T^{\mu\nu} \nabla_\mu \nabla_\nu \phi
+ b_5 R_{\mu \nu} T^{\mu \nu}
\nonumber \\
& +  b_6 RT + b_7 T +  c_1 G^{\mu \nu} \nabla_\mu \phi \nabla_\nu \phi +  c_2 R (\nabla
\phi)^2 +  c_3 R \square \phi
\nonumber \\
& +  d_1 R^2 +  d_2 R^{\mu \nu} R_{\mu \nu} +  d_3 \left( R^2 - 4 R^{\mu \nu} R_{\mu \nu} + R_{\mu \nu \sigma \rho} R^{\mu \nu \sigma \rho} \right)
\nonumber \\
&  +  d_4 \epsilon^{\mu \nu \lambda \rho} \tensor{C}{_{\mu
    \nu}^{\alpha \beta}} C_{\lambda \rho \alpha \beta} + e_1 T^{\mu
  \nu} T_{\mu \nu} + e_2 T^2
\bigg\}.
\label{eq:S1}
\end{align}
Here $C_{\mu \nu \lambda \rho}$ is the Weyl tensor and $\epsilon^{\mu
  \nu \lambda \rho}$ is the antisymmetric tensor (our conventions for
these are given in Appendix \ref{app:notation}).
There are additional terms with four
derivatives that one can write down,
but all such terms can be eliminated by integration by parts.
Finally, the metric ${\bar g}_{\mu\nu}$ which appears in
the matter action $S_{\rm m}$ in Eq.\ (\ref{eq:start}) is given
by\footnote{We call this metric the Jordan frame metric, in an
  extension of the usual terminology which applies to the case when
  the relation (\ref{eq:JordanMetric}) between the two metrics is just
  a conformal transformation.}
\begin{align}
{\bar g}_{\mu\nu} =&  e^{\alpha} g_{\mu\nu} + \epsilon e^\alpha \left[
     \beta_1 \nabla_\mu \phi \nabla_\nu \phi +
  \beta_2 (\nabla \phi)^2 g_{\mu \nu} +
     \beta_3 \square \phi g_{\mu \nu} +
 \beta_4 \nabla_\mu \nabla_\nu \phi +
\beta_5 R_{\mu\nu} +
\beta_6 R g_{\mu\nu}\right] \nonumber \\
&
+ O(\epsilon^2).
\label{eq:JordanMetric}
\end{align}
All of the coefficients $a_i, b_i,c_i, d_i, e_i$, $\beta_i$ and
$\alpha$ are
arbitrary functions of $\phi$.

Let us briefly discuss each of the perturbative terms. The terms with
coefficients $a_i$ are corrections to the kinetic term of the scalar
field. The $b_i$ and $\beta_i$ terms are couplings between the scalar field and the
stress-energy tensor, or between curvature and the stress-energy
tensor. The  $c_i$ terms are kinetic couplings
between the scalar field and gravity.
The $d_i$
terms are quadratic curvature terms, which we have chosen to write as an
$R^2$ term, an $R_{\mu \nu} R^{\mu\nu}$ term, and the Gauss-Bonnet
term. Any constant piece of the coefficient $d_3$ is a topological
term and may be
omitted. The term $d_4$ is the gravitational Chern-Simons term, which
may be excluded if one wishes to introduce parity as a symmetry of the
theory, and again, any constant component of $d_4$ is topological and
may be omitted. Finally, the $e_i$ terms are quadratic in the
stress-energy tensor.

Note that several of the terms in the action (\ref{eq:S1}) are ``higher
derivative'' terms, that is, they give rise to contributions to the
equations of motion containing derivatives of order three or higher.
The specific terms are those parameterized by the coefficients $a_3$,
$b_3, \ldots, b_6$, $c_2$, $c_3$, $d_1$, $d_2$ and $\beta_3, \ldots,
\beta_6$.  As discussed in the Introduction and in Appendix
\ref{app:backsubs}, we will choose to define our theory by treating
these terms perturbatively, which excludes the extra degrees
of freedom and instabilities that are normally associated with higher
derivative terms.

We also note that the theory (\ref{eq:start}) satisfies the weak
equivalence principle, to linear order in $\epsilon$, as we show in
Appendix \ref{app:WEP}.  That is, objects with negligible
self-gravity with different compositions all experience the same
acceleration.  It is not {\it a priori} obvious that the principle
should be satisfied since, as
we show in Appendix \ref{app:WEP}, violations of the principle
generically arise whenever the matter stress energy tensor appears
explicitly in the gravitational action, as in Eq.\ (\ref{eq:start}).

\ifx\doejects\undefined
\else
        \eject
\fi
\section{Transformation Properties of the Action}
\label{sec:techs}

The description of the theory provided by Eqs.\ (\ref{eq:start}) --
(\ref{eq:JordanMetric})
is very redundant, in part because of the freedom to perform field
redefinitions.  In this section we derive how the various
coefficients
in the action (\ref{eq:start}) are modified under various
transformations.
In the next section we will use these transformation laws
to derive a canonical representation of the theory, involving only
nine free functions.

\subsection{Expansion of the Matter Action}
\label{sec:techs:coupling}

Consider first the perturbative terms parameterized by $\beta_1,
\ldots, \beta_6$, in the definition
(\ref{eq:JordanMetric}) of the Jordan metric ${\bar g}_{\alpha\beta}$,
which appears in the matter action $S_{\rm m}[{\bar
  g}_{\alpha\beta},\psi_{\rm m}]$.  Using the definition (\ref{eq:Tabdef})
of the stress-energy tensor, we can eliminate these terms in favor of
terms in the action involving $T_{\alpha\beta}$.  Specifically we have
from Eq.\ (\ref{eq:Tabdef}) that
\begin{align}
S_{\rm m}[ e^\alpha (g_{\mu\nu}+ \delta
g_{\mu\nu}),\psi_{\rm m}] =
S_{\rm m}[ e^\alpha g_{\mu\nu},\psi_{\rm m}] +
\frac{1}{ 2} \int d^4 x \sqrt{-g} e^{2 \alpha} T^{\mu\nu} \delta
g_{\mu\nu} + O(\delta g^2).
\end{align}
Choosing
\begin{align}
\delta g_{\mu\nu} =  \epsilon [
     {\tilde \beta}_1 \nabla_\mu \phi \nabla_\nu
     \phi +
  {\tilde \beta}_2 (\nabla \phi)^2 g_{\mu \nu} +
     {\tilde \beta}_3 \square \phi g_{\mu \nu}
+ {\tilde \beta}_4 \nabla_\mu \nabla_\nu \phi
+ {\tilde \beta}_5 R_{\mu\nu}
+ {\tilde \beta}_6 R g_{\mu\nu}
]
\label{eq:tr00a}
\end{align}
then gives a transformation of the action (\ref{eq:start})
characterized by the following changes in the coefficients:
\begin{align}
\begin{array}{cc}
\delta \beta_1 = - {\tilde \beta}_1, \ \ \ \  & \delta b_1 = \frac{1}{2} e^{2
  \alpha} {\tilde \beta}_1, \\
\delta \beta_2 = - {\tilde \beta}_2, \ \ \ \  & \delta b_2 = \frac{1}{2} e^{2
  \alpha} {\tilde \beta}_2, \\
\delta \beta_3 = - {\tilde \beta}_3, \ \ \ \  & \delta b_3 = \frac{1}{2} e^{2
  \alpha} {\tilde \beta}_3, \\
\delta \beta_4 = - {\tilde \beta}_4, \ \ \ \  & \delta b_4 = \frac{1}{2} e^{2
  \alpha} {\tilde \beta}_4, \\
\delta \beta_5 = - {\tilde \beta}_5, \ \ \ \  & \delta b_5 = \frac{1}{2} e^{2
  \alpha} {\tilde \beta}_5, \\
\delta \beta_6 = - {\tilde \beta}_6, \ \ \ \  & \delta b_6 = \frac{1}{2} e^{2
  \alpha} {\tilde \beta}_6.
\end{array}
\label{eq:tr00}
\end{align}
Here the parameters ${\tilde \beta}_i$ can be arbitrary functions of
$\phi$.  Similarly choosing
$
\delta g_{\mu\nu} =  \epsilon {\tilde \alpha} g_{\mu\nu}
$
gives a transformation
characterized by
\begin{align}
\begin{array}{cc}
\delta \alpha = - \epsilon {\tilde \alpha}_, \ \ \ \  & \delta b_7 = \frac{1}{2} e^{2
  \alpha} {\tilde \alpha}.
\end{array}
\label{eq:tr01}
\end{align}

\subsection{Field Redefinitions Involving just the Scalar Field}
\label{sec:techs:scalar}

Consider a perturbative field redefinition of the form
\begin{align}
\phi = \psi + \epsilon \gamma,
\label{eq:redefine1}
\end{align}
where the quantity $\gamma$ can in general depend on any of the fields
and their derivatives.  To leading order in $\epsilon$, the change in
the action (\ref{eq:start}) is
then proportional
to the zeroth-order equation of motion (\ref{eq:eom0b}) for $\phi$.
Relabeling
$\psi$ as $\phi$, the change induced in the action is
\begin{align}
\delta S = \epsilon \int d^4 x \sqrt{-g} \gamma \left[ \square \phi
  - U' + \frac{1}{2} e^{2 \alpha} \alpha' T \right].
\label{eq:scalarchange}
\end{align}
There are three special cases that will be useful:

\begin{enumerate}

\item First, choose
\begin{align}
\phi = \psi + \epsilon \sigma_1 T,
\label{eq:ch1}
\end{align}
where $\sigma_1$ is an arbitrary function\footnote{Because we are working to linear order in $\epsilon$,
 it does not matter whether we take $\sigma_1$ to be a function of
 $\phi$ or of $\psi$.}
of $\psi$, and $T$ is the trace of the stress-energy tensor.
Substituting this into Eq.\ (\ref{eq:scalarchange})
and comparing with the general action (\ref{eq:S1}),
we find the following transformation law for the coefficients:
\begin{align}
\begin{array}{cc}
\delta b_3 = \sigma_1, \ \ \ \ &
\delta b_7 = - U' \sigma_1, \\
\delta e_2 = \frac{1}{2} \alpha' e^{2 \alpha} \sigma_1. &
\end{array}
\label{eq:tr0}
\end{align}

\item Second, we use the field redefinition
\begin{align}
\phi = \psi + \epsilon \sigma_2
[ \square \psi + U'(\psi)].
\label{eq:ch2}
\end{align}
Here the second term in the square
bracket is included in order to maintain canonical normalization of
the scalar field, that is, to avoid generating terms in the action of
the form $f(\phi) (\nabla \phi)^2$.
The resulting transformation law is
\begin{align}
\begin{array}{cc}
\delta a_3 = \sigma_2, \ \ \ \ &
\delta b_3 = \frac{1}{2} e^{2 \alpha}\alpha' \sigma_2, \\
\delta b_7 = \frac{1}{2} \alpha' e^{2 \alpha} U' \sigma_2, \ \ \ \  &
\delta U = \epsilon (U')^2 \sigma_2.
\end{array}
\label{eq:tr2}
\end{align}

\item Third, consider the field redefinition
\begin{align}
\phi = \psi + \epsilon \sigma_3
 - \epsilon \sigma_3^\prime (\nabla \psi)^2/U',
\label{eq:ch3}
\end{align}
where $\sigma_3$ is a function of $\psi$ and again the particular
combination of terms is chosen to maintain canonical normalization.
Substituting into Eq.\ (\ref{eq:scalarchange}), performing some
integrations by parts and comparing with Eq.\ (\ref{eq:S1})
gives the transformation law
\begin{align}
\begin{array}{cc}
\delta a_2 = -\sigma_3^\prime/U', \ \ \ \ &
\delta b_2 = -\frac{1}{2} e^{2 \alpha}\alpha' \sigma_3^\prime/U', \\
\delta b_7 = \frac{1}{2} e^{2 \alpha} \alpha'  \sigma_3, \ \ \ \  &
\delta U = \epsilon U' \sigma_3.
\end{array}
\label{eq:tr3}
\end{align}
Note that this transformation is not well defined in general in the
limit $U^\prime \to 0$, because of the factors of $1/U^\prime$.
However, it is well defined in the limit $U^\prime \to 0$,
$\sigma_3^\prime \to 0$ with $\sigma_3^\prime / U^\prime$ kept constant.
\end{enumerate}

\small
\begin{table}
\centering
\begin{tabular}{|c|c||c|c|c|c|c|c|c|c|c|c|c|}
\hline
Coeff. &Term & $\sigma_1$ &$\sigma_2$ & $\sigma_3$ & $\sigma_4$ & $\sigma_5$ & $\sigma_6$ & $\sigma_7$ & $\sigma_8$ & $\sigma_9$ &$\sigma_{10}$ &$\sigma_{11}$
\\ \hline
\hline
$a_1\ \ \ \ $ & $(\nabla \phi)^4$& &  & & & & $\star$ & &$\star$ & $\star$
& & \\
\hline
$a_2\ \ \ \ $ &$\square \phi (\nabla \phi)^2$ & &  &$\star$ & & & &$\star$ & &$\star$ &  & \\
\hline
$a_3\ \ \dagger \ $ &$(\square \phi)^2$ & & $\star$ & & & & & & & &  & \\
\hline
$b_1\ \ \ \ $ &$T^{\mu\nu} \nabla_\mu \phi \nabla_\nu \phi$ & &  & & & & & &$\star$  & &  & $\star$\\
\hline
$b_2\ \ \ \ $ & $T (\nabla \phi)^2$& &  &$\star$ & & &$\star$  & & & &$\star$  &$\star$ \\
\hline
$b_3\ \ \dagger \ $ &$T \square \phi$ &$\star$ & $\star$ & & & & &$\star$ & & &  & \\
\hline
$b_4\ \ \dagger \ $ &$T^{\mu\nu} \nabla_\mu \nabla_\nu \phi$ & &  & & & & & & & $\star$ &  & \\
\hline
$b_5\ \ \dagger \ $ &$R^{\mu\nu} T_{\mu\nu}$ & &  & & &$\star$ & & & & &  &$\star$ \\
\hline
$b_6\ \ \dagger \ $ &$R T$ & &  & &$\star$ & & & & & &$\star$  &$\star$ \\
\hline
$b_7\ \ \ \ $ &$T$ &$\star$
 &$\star$  &$\star$ & $\star$&$\star$ & $\star$&$\star$ &$\star$ &$\star$ &$\star$  & $\star$\\
\hline
$c_1\ \ \ \ $ &$G^{\mu\nu} \nabla_\mu \phi \nabla_\nu \phi$ & &  & & & $\star$ &  & &$\star$ &$\star$ &  & \\
\hline
$c_2\ \ \dagger \ $ &$R (\nabla \phi)^2$ & &  & & $\star$& &$\star$ & & & &  & \\
\hline
$c_3\ \ \dagger \ $ &$R \square \phi$ & &  & & & & &$\star$ & & &  & \\
\hline
$d_1\ \ \dagger \ $ &$R^2$ & &  & &$\star$ &$\star$ & & & & &  & \\
\hline
$d_2\ \ \dagger \ $ &$R^{\mu\nu} R_{\mu\nu} $ & &  & & &$\star$ & & &
& &  & \\
\hline
$d_3\ \ \ \ $ &Gauss-Bonnet & &  & & & & & & & &  & \\
\hline
$d_4\ \ \ \ $ &Chern-Simons & &  & & & & & & & &  & \\
\hline
$e_1\ \ \ \ $ &$T^{\mu\nu} T_{\mu\nu}$ & &  & & & & & & &&  & $\star$\\
\hline
$e_2\ \ \ \ $ &$T^2$ & $\star$ &  & & & & & & & & $\star$ & \\
\hline
&$U$\ (potential) & &$\star$  &$\star$ &$\star$ &$\star$ & $\star$&$\star$ &$\star$ &$\star$ &  & \\
\hline

\end{tabular}
\caption{This table shows which of the terms
  in our action
  (\protect{\ref{eq:S0}}) are
  affected by each of the eleven field redefinitions (\protect{\ref{eq:ch1}}) --
  (\protect{\ref{eq:ch11}}) that are parameterized by the functions
  $\sigma_1(\phi), \ldots, \sigma_{11}(\phi)$.  The columns represent the
  redefinitions, and the rows represent terms.  Daggers $\dagger$ in first column
indicate ``higher derivative'' terms, that is, terms that give
contributions to the equations of motion containing time derivatives of
higher than second order. Stars $\star$
  indicate that the coefficient of that row's term is altered by
  that column's field redefinition.  We omit the
coefficients $\alpha$ and $\beta_1, \ldots , \beta_6$
since those coefficients are degenerate with $b_1, \ldots, b_7$ by
Eqs.\ (\protect{\ref{eq:tr00}}) and (\protect{\ref{eq:tr01}}).}
\label{tab:transformations}
\end{table}
\normalsize

\subsection{Field Redefinitions Involving the Metric}
\label{sec:techs:metric}

We now consider a more general class of field redefinitions, where in
addition to redefining the scalar field via Eq.\ (\ref{eq:redefine1}),
we also perturbatively redefine the metric via
\begin{align}
g_{\alpha\beta} = {\hat g}_{\alpha\beta} + \epsilon F_{\alpha\beta}.
\label{eq:chm}
\end{align}
Here the quantity $F_{\alpha\beta}$ can depend on $\psi$, ${\hat
  g}_{\alpha\beta}$, their derivatives and the stress energy tensor.
The corresponding change in the action is proportional to the
equation of motion (\ref{eq:eom0a}).  Relabeling ${\hat
  g}_{\alpha\beta}$ as $g_{\alpha\beta}$ and $\psi$ as $\phi$, the
total change in the action is
\begin{align}
\delta S =& \frac{\epsilon}{2} \int d^4 x \sqrt{-g} F_{\alpha\beta}
\left[ - \mpl^2 G^{\alpha\beta} + \nabla^\alpha\phi \nabla^\beta \phi
- \frac{1}{2} (\nabla \phi)^2 g^{\alpha\beta} - U g^{\alpha\beta}
+ e^{2 \alpha} T^{\alpha\beta}\right] \nonumber \\
& +
\epsilon \int d^4 x \sqrt{-g} \gamma \left[ \square \phi
  - U' + \frac{1}{2} e^{2 \alpha} \alpha' T \right].
\label{eq:tensorchange}
\end{align}
Note that this formula includes the effect of the change in the Jordan
frame metric (\ref{eq:JordanMetric}) caused by the transformation
(\ref{eq:chm}).
We now consider seven different transformations of this type:

\begin{enumerate}
\setcounter{enumi}{3}
\item The first case is a change to the metric proportional to $R
g_{\alpha\beta}$.  In order to maintain canonical normalization of
both the metric and the scalar field, that is, to avoid terms of the
form $f(\phi) (\nabla \phi)^2$ and $f(\phi) R$, we need the following
combination of terms in the field redefinition:
\bes
\begin{align}
g_{\alpha\beta} =& {\hat g}_{\alpha\beta} - 2 \epsilon \sigma_4^\prime \left(
  \frac{\mpl^2}{U} {\hat R} + 4  \right) {\hat g}_{\alpha\beta},  \\
\phi =& \psi + 4 \epsilon \sigma_4,
\end{align}
\label{eq:ch4}
\ees
for some function $\sigma_4(\psi)$.  Substituting into Eq.\
(\ref{eq:tensorchange}),
performing some integrations by parts and comparing with Eq.\
(\ref{eq:S1}) we obtain for the transformation law
\begin{align}
\begin{array}{cc}
\delta b_7 = 2 e^{2 \alpha} \alpha' \sigma_4- 4 e^{2 \alpha} \sigma_4^\prime, \ \ \ \ &
\delta c_2 = \frac{\mpl^2}{U}  \sigma_4^\prime, \\
\delta d_1 = -\frac{\mpl^4}{U}  \sigma_4^\prime, \ \ \ \  &
\delta b_6 = - \frac{e^{2 \alpha}}{U} \mpl^2 \sigma_4^\prime, \\
\delta U = 4 \epsilon \left[U' \sigma_4  - 4 U \sigma_4^\prime \right].
\end{array}
\label{eq:tr4}
\end{align}

\item Next consider changes to the metric proportional to
  $R_{\alpha\beta}$.  In order to maintain canonical normalizations
we use the following
combination of terms in the field redefinition:
\bes
\begin{align}
g_{\alpha\beta} =& {\hat g}_{\alpha\beta}(1 - 2 \epsilon
\sigma_5^\prime)
- 2 \epsilon \frac{\mpl^2}{U} \sigma_5^\prime {\hat R}_{\alpha\beta},  \\
\phi =& \psi +  \epsilon \sigma_5,
\end{align}
\label{eq:ch5}
\ees
for some function $\sigma_5(\psi)$.  This gives
the transformation law
\begin{align}
\begin{array}{cc}
\delta b_7 = \frac{1}{2} e^{2 \alpha} \alpha' \sigma_5- e^{2 \alpha} \sigma_5^\prime, \ \ \ \ &
\delta c_1 = -\frac{\mpl^2}{U}  \sigma_5^\prime, \\
\delta d_1 = -\frac{\mpl^4}{2 U}  \sigma_5^\prime, \ \ \ \  &
\delta d_2 = \frac{\mpl^4}{U} \sigma_5^\prime,\\
\delta b_5 = - \frac{\mpl^2}{U} e^{2 \alpha} \sigma_5^\prime, \ \ \ \ &
\delta U = \epsilon \left[U' \sigma_5  - 4 U \sigma_5^\prime \right].
\end{array}
\label{eq:tr5}
\end{align}

\item The next case is a change to the metric proportional to $(\nabla
  \phi)^2 g_{\alpha\beta}$.
To maintain canonical normalization of the scalar field, we need in
addition a change to the scalar field, with the combined
transformation being
\bes
\begin{align}
g_{\alpha\beta} =& {\hat g}_{\alpha\beta} - 2
\epsilon  \frac{\sigma_6^\prime}{U} ({\hat \nabla} \psi)^2 {\hat
  g}_{\alpha\beta},  \\
\phi =& \psi +  4 \epsilon \sigma_6,
\end{align}
\label{eq:ch6}
\ees
for some function $\sigma_6$.  The resulting
transformation law for the coefficients is
\begin{align}
\begin{array}{cc}
\delta a_1 = \sigma_6^\prime/U, \ \ \ \ &
\delta b_2 = -e^{2 \alpha} \sigma_6^\prime/U, \\
\delta b_7 = 2 e^{2 \alpha} \alpha' \sigma_6, \ \ \ \  &
\delta c_2 = - \sigma_6^\prime \mpl^2/U,\\
\delta U = 4 \epsilon U' \sigma_6.
\end{array}
\label{eq:tr6}
\end{align}

\item Next consider changes to the metric proportional to
  $\square \phi g_{\alpha\beta}$.  The required form of field
  redefinition that preserves canonical normalization of $\phi$ is
\bes
\begin{align}
g_{\alpha\beta} =& {\hat g}_{\alpha\beta}
+ 2 \epsilon  \sigma_7 {\hat \square} \psi {\hat g}_{\alpha\beta},  \\
\phi =& \psi +  4 \epsilon U \sigma_7,
\end{align}
\label{eq:ch7}
\ees
for some function $\sigma_7$.  The coefficients in the action then
change according to
\begin{align}
\begin{array}{cc}
\delta a_2 = -\sigma_7, \ \ \ \ &
\delta b_3 = e^{2 \alpha} \sigma_7, \\
\delta b_7 = 2 e^{2 \alpha} \alpha' U \sigma_7, \ \ \ \  &
\delta c_3 = \mpl^2 \sigma_7,\\
\delta U = 4 \epsilon U U' \sigma_7.
\end{array}
\label{eq:tr7}
\end{align}

\item The fifth case is a change to the metric proportional to
  $\nabla_\alpha \phi \nabla_\beta \phi$.  The required form of field
  redefinition that preserves canonical normalization of $\phi$ is
\bes
\begin{align}
g_{\alpha\beta} =& {\hat g}_{\alpha\beta}
- 2 \epsilon  \frac{\sigma_8^\prime}{U}{\hat \nabla}_\alpha \psi {\hat \nabla}_\beta \psi,  \\
\phi =& \psi +  \epsilon  \sigma_8,
\end{align}
\label{eq:ch8}
\ees
for some function $\sigma_8$.  The coefficients in the action then
change according to
\begin{align}
\begin{array}{cc}
\delta a_1 = -\sigma_8^\prime/(2 U), \ \ \ \ &
\delta b_1 = -e^{2 \alpha} \sigma_8^\prime/U, \\
\delta b_7 = \frac{1}{2} e^{2 \alpha} \alpha'  \sigma_8, \ \ \ \  &
\delta c_1 = \mpl^2 \sigma_8^\prime/U,\\
\delta U =  \epsilon U' \sigma_8.
\end{array}
\label{eq:tr8}
\end{align}

\item Next consider a change in the metric
proportional to
  $\nabla_\alpha \nabla_\beta \phi$.  To preserve canonical
  normalization of $\phi$ we use the redefinitions
\bes
\begin{align}
g_{\alpha\beta} =& {\hat g}_{\alpha\beta}
+ 2 \epsilon \sigma_9 {\hat \nabla}_\alpha {\hat \nabla}_\beta \psi,  \\
\phi =& \psi +  \epsilon U \sigma_9 ,
\end{align}
\label{eq:ch9}
\ees
for some function $\sigma_8$.  The coefficients in the action then
change according to
\begin{align}
\begin{array}{cc}
\delta a_1 = -\frac{1}{2} \sigma_9^\prime, \ \ \ \ &
\delta a_2 = -\sigma_9, \\
\delta b_4 = e^{2 \alpha} \sigma_9, \ \ \ \ &
\delta b_7 = \frac{1}{2} e^{2 \alpha} \alpha' U \sigma_9, \\
\delta c_1 = \mpl^2 \sigma_9^\prime, \ \ \ \ &
\delta U =  \epsilon U U' \sigma_9.
\end{array}
\label{eq:tr9}
\end{align}

\item A simple case is when the change in the metric is proportional
  to $T g_{\alpha\beta}$, for which no change to the scalar field is
  required.  The redefinition is
\begin{align}
g_{\alpha\beta} =& {\hat g}_{\alpha\beta}
+ 2 \epsilon \sigma_{10} T {\hat g}_{\alpha\beta},
\label{eq:ch10}
\end{align}
for some function $\sigma_{10}$.  The transformation law for the
coefficients is
\begin{align}
\begin{array}{cc}
\delta b_2 = -\sigma_{10}, \ \ \ \ &
\delta b_7 = -4 \sigma_{10} U, \\
\delta e_2 = e^{2 \alpha} \sigma_{10}, \ \ \ \ &
\delta b_6 = \mpl^2 \sigma_{10}.
\end{array}
\label{eq:tr10}
\end{align}

\item Similarly, no transformation to the scalar is required for the
  case of a change in the metric proportional to $T_{\alpha\beta}$.
The redefinition is
\begin{align}
g_{\alpha\beta} =& {\hat g}_{\alpha\beta}
+ 2 \epsilon \sigma_{11} T_{\alpha\beta},
\label{eq:ch11}
\end{align}
for some function $\sigma_{11}$, and the corresponding transformation
law is
\begin{align}
\begin{array}{cc}
\delta b_1 = \sigma_{11}, \ \ \ \ &
\delta b_2 = -\frac{1}{2} \sigma_{11}, \\
\delta b_7 = - \sigma_{11} U, \ \ \ \ &
\delta e_1 = e^{2 \alpha} \sigma_{11}, \\
\delta b_5 = -\mpl^2 \sigma_{11}, \ \ \ \ &
\delta b_6 = \frac{1}{2} \mpl^2 \sigma_{11}.
\end{array}
\label{eq:tr11}
\end{align}

\end{enumerate}

The eleven\footnote{We could also consider a twelfth redefinition
  given by $g_{\alpha\beta} = {\hat g}_{\alpha\beta}(1 - 2
  \epsilon \sigma_{12}^\prime)$, $\phi = \psi + \epsilon \sigma_{12} - \epsilon \mpl^2
  \sigma_{12}^\prime {\hat R} / U'$.  However this redefinition is not
  independent of the first eleven; the same effect can be achieved by
  choosing $\sigma_1 = - e^{2\alpha} \sigma_{12}^\prime/U'$, $\sigma_3
  = -\sigma_{12}$, $\sigma_7 = \sigma_{12}^\prime/U'$, $\sigma_{10} =
  e^{2 \alpha} \alpha' \sigma_{12}^\prime/(2 U')$.}
field redefinitions (\ref{eq:ch1}) -- (\ref{eq:ch11}) are
summarized in
Table \ref{tab:transformations}, which shows which coefficients
are modified by which transformations.

\ifx\doejects\undefined
\else
        \eject
\fi
\section{Canonical Form of Action}
\label{sec:details}

In this section, we derive our final, reduced action
(\ref{eq:details:finalaction0})
from the starting action (\ref{eq:start}), using the transformation
laws derived in Sec.\ \ref{sec:techs}.
There is some freedom in which terms we choose to eliminate and which
terms we choose to retain.  We choose to eliminate all terms that give
higher derivatives in the equations of motion, so that the final
theory is not a ``higher derivative'' theory.  However, even after
this has been accomplished, there is still some freedom in how the
final
theory is represented.  We discuss this further in Sec.\
\ref{sec:details:final} below.
The order of operations in the derivation is important, since we
need to take care that terms which we have already set to zero
are not reintroduced by subsequent transformations.
Table \ref{tab:operations} summarizes our calculations and their
effects on the coefficients in the action at each stage in the
computation.

\subsection{Derivation}
\label{sec:derivation1}
The steps in the derivation are as follows:

\begin{table}
\centering
\footnotesize
\begin{tabular}{|c|c||c|c|c|c|c|c|c|c|c|c|}
\hline
Step &  & 1 & 2 & 3 & 3 & 4 &
5 &
6 &
6 &
7 &
Final
\\
% new stuff
\hline
 & Transformations  & ${\tilde \beta}_j$ & $\sigma_4,\sigma_5$ &
$\sigma_9$ & $\sigma_{10},\sigma_{11}$ &
$\sigma_6,\sigma_7$ &
$\sigma_2,\sigma_3$ &
$\sigma_8$ &
$\sigma_1$ &
${\tilde \alpha}$ &
\\
\hline
\hline
Coeff. &Term in Action&  & &  & &  &  &  &  & &
\\ \hline
$a_1\ \ \ \ $ & $(\nabla \phi)^4$& &  &$\star$ & & $\star$ & &$\star$  &
& & \checkmark \\
\hline
$a_2\ \ \ \ $ &$\square \phi (\nabla \phi)^2$ &  & &$\star$ &
&$\star$ & $\rightarrow 0$ & & & & $\circ$\\
\hline
$a_3\ \ \dagger \ $ &$(\square \phi)^2$ & & & & & & $\rightarrow 0$ & & &  & \\
\hline
$b_1\ \ \ \ $ &$T^{\mu\nu} \nabla_\mu \phi \nabla_\nu \phi$ & $\star$
& & &$\star$ & & & $\rightarrow 0$   & &  &$\circ$ \\
\hline
$b_2\ \ \ \ $ & $T (\nabla \phi)^2$&$\star$ &  &&$\star$  & $\star$&
$\star$ & & &  & \checkmark\\
\hline
$b_3\ \ \dagger \ $ &$T \square \phi$ &$\star$ &  & & &$\star$ &  $\star$&
&$\rightarrow 0$ &  & \\
\hline
$b_4\ \ \dagger \ $ &$T^{\mu\nu} \nabla_\mu \nabla_\nu \phi$ & $\star$ &
&$\rightarrow 0$ & & & & &  &  & \\
\hline
$b_5\ \ \dagger \ $ &$R^{\mu\nu} T_{\mu\nu}$ &$\star$ &$\star$ &
&$\rightarrow 0$ & & & & &  & \\
\hline
$b_6\ \ \dagger \ $ &$R T$ &$\star$ &$\star$  & &$\rightarrow 0$ & & & & &  & \\
\hline
$b_7\ \ \ \ $ &$T$ &
 &$\star$  &$\star$ & $\star$&$\star$ & $\star$&$\star$ &$\star$
 &$\rightarrow 0$  & \\
\hline
$c_1\ \ \ \ $ &$G^{\mu\nu} \nabla_\mu \phi \nabla_\nu \phi$ &
&$\star$&$\star$ &  && &$\star$    & &  &\checkmark \\
\hline
$c_2\ \ \dagger \ $ &$R (\nabla \phi)^2$ & & $\star$ & & &$\rightarrow 0$ &  & & &  & \\
\hline
$c_3\ \ \dagger \ $ &$R \square \phi$ & &  & & &$\rightarrow 0$ & & & & &   \\
\hline
$d_1\ \ \dagger \ $ &$R^2$ & &$\rightarrow 0$  & & & & & & & &   \\
\hline
$d_2\ \ \dagger \ $ &$R^{\mu\nu} R_{\mu\nu}$ & & $\rightarrow 0$ &  & & & &
& &  &  \\
\hline
$d_3\ \ \ \ $ &Gauss-Bonnet & &  & & & & & & & &  \checkmark \\
\hline
$d_4\ \ \ \ $ &Chern-Simons & &  & & & & & & & & \checkmark\\
\hline
$e_1\ \ \ \ $ &$T^{\mu\nu} T_{\mu\nu}$ & &  &  &$\star$ & & & &&  &
\checkmark \\
\hline
$e_2\ \ \ \ $ &$T^2$ &  &  & &$\star$ & & & &$\star$ &  &\checkmark \\
\hline
&$U$\ (potential) & &$\star$  &$\star$ & &$\star$ &
$\star$&$\star$ & & & \checkmark  \\
\hline
Coeff. &Term in ${\bar g}_{\mu\nu}$&   &  & &  &  &  &  & & &
\\ \hline
$\beta_1\ \ \ \ $ &$\nabla_\mu \phi \nabla_\nu \phi$ &
$\rightarrow 0$
  & & & & & & & &  & \\
\hline
$\beta_2\ \ \ \ $ &$(\nabla \phi)^2 g_{\mu\nu}$ &$\rightarrow 0$ & & & & & & & &  & \\
\hline
$\beta_3\ \ \dagger \ $ &$\square \phi g_{\mu\nu}$ &$\rightarrow 0$ &   & & & & & & &  & \\
\hline
$\beta_4\ \ \dagger \ $ &$\nabla_\mu \nabla_\nu \phi$ & $\rightarrow 0$  & & & & & & & &  & \\
\hline
$\beta_5\ \ \dagger \ $ &$R_{\mu\nu}$ &$\rightarrow 0$ &  & & & & & &  &  & \\
\hline
$\beta_6\ \ \dagger \ $ &$R g_{\mu\nu}$ &$\rightarrow 0$ &  & & & & &  & &  & \\
\hline
 & $\alpha$ (conf. factor) & &  & & & & & &  & $\star$ &\checkmark \\
\hline
\end{tabular}
\caption{
%\small
This table shows the progression of manipulations we make in
  this section. The second column on the left lists the various terms in the
  action (\protect{\ref{eq:S1}}), or in the Jordan-frame metric
  (\protect{\ref{eq:JordanMetric}}). The first column lists the
  corresponding coefficients; daggers $\dagger$ indicate higher
  derivative terms.  The numbers in the first row along the top refer
  to the numbered steps in the derivation in Sec.\
  \protect{\ref{sec:derivation1}}.
  The second row shows which transformation functions are used in
  each step.
  In the table, a star $\star$ indicates that the corresponding row's term
  receives a
  contribution from the corresponding column's reduction process,
  while $\rightarrow 0$
  indicates that the term has been eliminated.  The check marks
  $\checkmark$ in the
last column indicate the remaining terms that are non-zero in the
final action (\protect{\ref{eq:details:finalaction}}).  Finally,
circles $\circ$ in the last column indicate terms that are nonzero in
alternative forms of the final action obtained using the
transformations (\protect{\ref{eq:ch3}}) or (\protect{\ref{eq:ch8}}),
as discussed in Sec.\ \protect{\ref{sec:details:final}}.
%\normalsize
}
\label{tab:operations}
\end{table}

\normalsize

\begin{enumerate}

\item {\it Elimination of Derivative Terms in the Jordan Frame Metric:
  }
The transformation (\ref{eq:tr00}) can be used to eliminate all of the
terms involving derivatives in the Jordan frame metric
(\ref{eq:JordanMetric}), which are parameterized by the coefficients
$\beta_1, \ldots, \beta_6$.  This changes the coefficients of the
terms in the action that depend linearly on the stress energy tensor,
namely $b_1, \ldots, b_6$.
As discussed in Appendix \ref{app:WEP}, these terms involving the
stress-energy tensor
look like they might violate the weak equivalence principle, but in
fact they do not.

\item {\it Elimination of Higher Derivative, Quadratic in
  Curvature Terms: }
We next consider the terms in the action that are quadratic functions
of curvature, whose coefficients are $d_1$, $d_2$, $d_3$ and $d_4$.
The Chern-Simons term ($d_4$) and the Gauss-Bonnet term ($d_3$)
give rise to well behaved equations of motion (in the sense that they
not increase the number of degrees of freedom), so we do not attempt
to eliminate these terms.  By contrast, the terms proportional to the
squares of the Ricci scalar and Ricci tensor, parameterized by $d_1$
and $d_2$, do increase the number of degrees of freedom.
We can eliminate these terms by using the transformations
(\ref{eq:ch4}) and (\ref{eq:ch5}), with parameters chosen to be
\begin{align}
\sigma_4 = \int d\phi \, \frac{U}{\mpl^4} (d_1 + d_2/2), \ \
\ \ \ \sigma_5 = -\int d\phi \, \frac{U}{\mpl^4} d_2.
\end{align}
These transformations will then modify the coefficients $b_5$, $b_6$,
$b_7$, $c_1$ and $c_2$, as well as the potential $U$ (see Table \ref{tab:transformations}).

\item {\it Elimination of some of the Linear Stress-Energy
  Terms: }We next turn to
terms which depend linearly on the stress-energy
tensor, parameterized by $b_1, \ldots, b_6$.
First, we can eliminate the term $b_4 T^{\mu\nu} \nabla_\mu \nabla_\nu
\phi$ by using the transformation (\ref{eq:ch9}) with $\sigma_9 = -
e^{-2 \alpha} b_4$.  This gives rise to changes in the coefficients
$a_1$, $a_2$, $b_7$, $c_1$ as well as to the potential $U$.
Second, we can eliminate the terms parameterized by $b_5$ and $b_6$ by
using the transformations (\ref{eq:ch10}) and (\ref{eq:ch11}) with
the parameters $\sigma_{10} = - (b_6 + b_5/2)/\mpl^2$,
 $\sigma_{11} = b_5/\mpl^2$.  This changes the coefficients $b_1$,
 $b_2$, $b_7$, $e_1$ and $e_2$.

\item {\it Elimination of Kinetic Coupling of the Scalar to
  Curvature: } We next focus on the terms which kinetically couple the
scalar field to
gravity, namely $G^{\mu\nu} \nabla_\mu \phi \nabla_\nu \phi$, $R
(\nabla \phi)^2$ and
$R \square \phi$.  The first of these does not produce higher
derivative terms in the equation of motion, so we focus on the
remaining two terms, which are parameterized by
$c_2$ and $c_3$. These terms can be eliminated using the
transformations
(\ref{eq:ch6}) and (\ref{eq:ch7}), with the parameters chosen to be
\begin{align}
\sigma_6 = \int d\phi \, \frac{U}{\mpl^2} c_2, \ \ \ \ \ \sigma_7 = -
\frac{c_3}{\mpl^2}.
\end{align}
These transformations then give rise to changes in the coefficients
$a_1$, $a_2$, $b_2$, $b_3$, $b_7$ as well as to the potential $U$.

\item {\it Elimination of some of the Corrections to Scalar Field
    Kinetic Term: } Our action includes three corrections to the
  scalar kinetic
term, parameterized by $a_1$, $a_2$ and $a_3$.
Of these, only term $a_3$ contributes higher order derivatives to the
equations of motion.  We eliminate this term, and also the term
$a_2$, by using the transformations (\ref{eq:ch2}) and (\ref{eq:ch3})
with
\begin{align}
\sigma_2 = - a_3, \ \ \ \ \ \sigma_3= \int d\phi \, U' a_2.
\end{align}
This gives rise to corrections to the coefficients $b_2$, $b_3$ and
$b_7$ and to the potential $U$.

\item {\it Elimination of some Kinetic Couplings of the Scalar to
  Stress-Energy: } We next turn to the term $b_1 T^{\mu\nu} \nabla_\mu
\phi \nabla_\nu
\phi$.  We can eliminate this using the transformation (\ref{eq:ch8})
with the parameter choice
\begin{align}
\sigma_8 = \int d\phi \, b_1 U e^{-2 \alpha}.
\end{align}
This gives rise to changes in the coefficients $a_1$, $b_7$, $c_1$ and
$U$, from Table \ref{tab:transformations}.
We can also eliminate the term $b_3 T \square\phi$ by using the
transformation (\ref{eq:ch1}) with $\sigma_1 = -
 b_3$.  This changes the coefficients $e_2$ and
$b_7$.

\item {\it Elimination of Trace of Stress-Energy Tensor Term: }
The last step is to re-express the term $b_7 T$ in terms of an
$O(\epsilon)$ correction to the conformal factor $e^\alpha$ by using
the transformation (\ref{eq:tr01}) with ${\tilde \alpha} = - 2 e^{-2
  \alpha} b_7$.

\end{enumerate}

\subsection{Canonical Form of Action and Discussion}
\label{sec:details:final}

Applying the parameter specializations derived above to the action
(\ref{eq:start}) we arrive at our final result:
\begin{align}
S =&  \int d^4 x \sqrt{-g}
\left\{ \frac{\mpl^2}{2} R - \frac{1}{2}
  (\nabla \phi)^2 - U(\phi)\right\}
+ S_{\rm m}[e^{\alpha(\phi)} g_{\alpha\beta},\psi_{\rm m}] \nonumber \\
&
+ \epsilon \int d^4 x \sqrt{-g}  \bigg\{a_1 (\nabla \phi)^4
+  b_2 T (\nabla \phi)^2
+  c_1 G^{\mu \nu} \nabla_\mu \phi \nabla_\nu \phi
\nonumber \\
& +  d_3 \left( R^2 - 4 R^{\mu \nu} R_{\mu \nu} + R_{\mu \nu \sigma \rho} R^{\mu \nu \sigma \rho} \right)
%\nonumber \\ &
  +  d_4 \epsilon^{\mu \nu \lambda \rho} \tensor{C}{_{\mu
    \nu}^{\alpha \beta}} C_{\lambda \rho \alpha \beta} + e_1 T^{\mu
  \nu} T_{\mu \nu} + e_2 T^2
\bigg\}.
\label{eq:details:finalaction}
\end{align}
This action contains nine free functions of $\phi$: $U, \alpha, a_1, b_2,
c_1, d_3, d_4, e_1, e_2$.  The corresponding equations of motion do
not contain any higher derivative terms and are presented in Appendix \ref{app:eoms}.

Our final result  (\ref{eq:details:finalaction})
can be re-expressed in a number of equivalent forms:

\begin{itemize}
\item First, the term $b_2 T (\nabla \phi)^2$ in the action can be
eliminated in favor of a term proportional to $e^{2 \alpha} \beta_2
(\nabla \phi)^2 g_{\mu\nu}$ in the Jordan frame metric
(\ref{eq:JordanMetric}) using the transformation (\ref{eq:tr00}).  As
discussed in Appendix \ref{app:WEP} the latter representation makes
explicit that the weak equivalence principle is satisfied.

\item The term $b_2 T (\nabla \phi)^2$ could also be eliminated in
  favor of a term $a_2 \square \phi (\nabla \phi)^2$, using the
  transformation (\ref{eq:ch3}) parameterized by $\sigma_3$, as long as $\alpha' \ne
  0$\footnote{More precisely the criterion is that the zeroth order
    term in the expansion in $\alpha'$ in powers of $\epsilon$ is
    nonzero.  A nonzero $\alpha'$ that is proportional to $\epsilon$
    would be insufficient to allow this transformation.}.
The dynamics of a scalar quintessence field with kinetic terms of the
latter type have recently been explored in detail in Ref.\
\cite{Deffayet:2010qz}, who called the mixing of the scalar and metric
kinetic terms in the equations of motion ``kinetic braiding''.
The representation of this term as $a_2 \square \phi (\nabla \phi)^2$
has some advantages for cosmological analyses: in this representation
the dynamics of the term are confined to the scalar sector, while in
the $b_2$ representation they are coupled to matter.

\item The term $a_1 (\nabla \phi)^4$ can be eliminated in favor
of a term $b_1 T^{\mu\nu} \nabla_\mu \phi \nabla_\nu \phi$, using the
transformation (\ref{eq:ch8}) parameterized by $\sigma_8$.

\item Alternatively, the term $c_1 G^{\mu\nu} \nabla_\mu \phi
  \nabla_\nu \phi$ can be eliminated in favor of a term
$b_1 T^{\mu\nu} \nabla_\mu \phi \nabla_\nu \phi$, using the
transformation (\ref{eq:ch8}) parameterized by $\sigma_8$.  Our result
in this
representation agrees with that of Weinberg \cite{Weinberg2008} when
all the matter terms are dropped.
The $c_1$ representation has the advantage over the $b_1$
representation that the corrections are confined to the scalar sector
and do not involve matter.  The term $c_1 G^{\mu\nu} \nabla_\mu \phi
\nabla_\nu \phi$ has interesting effects: it can give rise to a
self-tuning cosmology as well as potentially support a Vainshtein
screening mechanism \cite{Charmousis2011}.

\item As discussed in Appendix \ref{app:WEP}, it is possible to
  eliminate all the stress-energy terms from the action by applying
field redefinitions.  This yields a form of the theory in which the
weak equivalence principle is manifest.  However, the resulting
action contains higher derivative terms,
unlike all the representations discussed so far in this subsection.
As discussed in the Introduction and in Appendix \ref{app:backsubs}, to
define the theory when higher derivative terms are present we use
the reduction of order technique applied to the equations of motion.

\item Finally, the result can be cast in the Jordan conformal frame by
  doing a conformal transformation, followed by some field
  redefinitions to simplify the answer.  The result is similar in form
  to the Einstein frame action   (\ref{eq:details:finalaction}):
\begin{align}
S[{\tilde g}_{\alpha\beta},{\tilde \phi},\psi_{\rm m}] =&  \int d^4 x \sqrt{-{\tilde g}}
\left[ \frac{1}{2} \mpl^2 e^{-\alpha} {\tilde R} - \frac{1}{2}
  ({\tilde \nabla} {\tilde \phi})^2 - {\tilde U}({\tilde \phi})\right]
+ S_{\rm m}[{\tilde g}_{\alpha\beta},\psi_{\rm m}] \nonumber \\
&
+ \epsilon \int d^4 x \sqrt{-g}  \bigg\{{\tilde a}_1 ({\tilde \nabla}
{\tilde \phi})^4
+  {\tilde b}_2 T ({\tilde \nabla} {\tilde \phi})^2
+  {\tilde c}_1 {\tilde G}^{\mu \nu} {\tilde \nabla}_\mu {\tilde \phi}
{\tilde \nabla}_\nu {\tilde \phi}
\nonumber \\
& +  {\tilde d}_3 \left( {\tilde R}^2 - 4 {\tilde R}^{\mu \nu}
  {\tilde R}_{\mu \nu} + {\tilde R}_{\mu \nu \sigma \rho} {\tilde R}^{\mu \nu \sigma \rho} \right)
%\nonumber \\ &
  +  {\tilde d}_4 \epsilon^{\mu \nu \lambda \rho} \tensor{{\tilde C}}{_{\mu
    \nu}^{\alpha \beta}} {\tilde C}_{\lambda \rho \alpha \beta}
\nonumber \\ &
+
{\tilde e}_1 T^{\mu
  \nu} T_{\mu \nu} + {\tilde e}_2 T^2
\bigg\}.
\label{eq:details:finalactionJ}
\end{align}
Here ${\tilde g}_{\mu\nu} = e^\alpha g_{\mu\nu}$ and the field
${\tilde \phi}$ is a function of $\phi$, where the function is chosen
to give canonical normalization for ${\tilde \phi}$ in the Jordan
frame action (\ref{eq:details:finalactionJ}).
All of the functions ${\tilde U}$, ${\tilde a}_1$, etc. in this action
differ from
the corresponding functions in the Einstein frame representation
(\ref{eq:details:finalaction}), but can in principle be computed in
terms of them.
The Jordan frame result (\ref{eq:details:finalactionJ})
can also be cast in a number of different forms using linearized field
redefinitions, just as for the Einstein frame result
(\ref{eq:details:finalaction}).
Note that the stress energy tensor we use is the same
in both frames, and is defined in Appendix \ref{app:notation}.
The result (\ref{eq:details:finalactionJ}) matches that found by Park
\textit{et al.}\ \cite{Watson2010} (up to some minor adjustments, see
Appendix \ref{app:compare}).

\end{itemize}

We note that the Chern-Simons term ($d_4$) gives rise to third-order
derivatives in the equations of motion [see Eqs.\ \eqref{eq:tensoreom}
and \eqref{eq:chernsimons} below]. However, with the choice of
foliation\footnote{This choice requires the assumption that $\nabla
\phi$ is timelike everywhere, which will be true in cosmological
applications
when perturbations are sufficiently small.} given by surfaces of
constant $\phi$, there are no
third-order time derivatives, and so the Chern-Simons term is not a
higher derivative term according to our definition (see the discussion
in Sec.\ \ref{sec:approach} above), and is not subject to the
Ostrogradski instability.  For further
discussion of the Chern-Simons term in gravitational theories, see,
e.g., Ref.\ \cite{Yunes2009}. As a parity-violating term, this term
modifies the propagation speed of different polarizations of
gravitons.

In the above derivation, we eliminated higher derivative terms using
field redefinitions.  As discussed in Appendix \ref{app:backsubs}, an
alternative but equivalent procedure is to derive a form of the
action which explicitly exhibits the extra degrees of freedom
associated with the higher derivative terms, and then integrate out
those degrees of freedom at tree level.  This is shown explicitly for
higher derivatives of the scalar field in Appendix \ref{app:backsubs},
and can also be shown explicitly for the terms
$d_1$ and $d_2$ involving higher derivatives of the metric.
A third, equivalent method is to perform a reduction of order
procedure at the level of the equations of motion, as discussed in
the Introduction and in Appendix \ref{app:backsubs}.

The above derivation confirms the general argument made in the
Introduction that it should not be necessary to include
higher derivative terms in the action.
This is because the new terms that are generated when one
eliminates the higher derivative terms should
already be included in the derivative expansion.
In the above derivation, if we eliminate from the start
the higher derivative terms ($a_3, b_3, b_4, b_5, b_6, c_2, c_3, d_1,
d_2$), then we must also forbid all transformations
that generate these terms, which includes all the transformations we
have considered except Eqs.\ (\ref{eq:tr00}), (\ref{eq:tr01}),
(\ref{eq:ch3}) and (\ref{eq:ch8}).
The above derivation
gets modified by dropping steps 2, 3, and 4, the portion of step 5 that
sets
$a_3$ to zero, and the portion of step 6 that sets $b_3$ to zero.
The final result (\ref{eq:details:finalaction}) is unchanged.

In a similar vein, the correct result can also be obtained by
omitting from the initial action all terms involving the stress
energy tensor, that is, the terms parameterized by
$b_1$, \ldots, $b_7$ and $e_1$, $e_2$.  If one follows all the steps
of the derivation in Table \ref{tab:operations},
the same final result is obtained, and all the final coefficients are
nonzero in general.  This occurs because all the terms involving the
stress energy tensor have alternative representations not involving it
(although they do involve higher derivatives).  Thus, from this point
of view, it is not necessary to include in the action stress-energy
terms.

However, it is {\it not} possible to do without both the higher
derivative terms and the stress-energy terms.  Suppose we
throw out at the start all the higher
derivative terms in both the action (\ref{eq:start})
and Jordan frame metric (\ref{eq:JordanMetric}), and in addition
omit all the stress-energy terms in the action.
This would yield a version of the
action (\ref{eq:start}) involving only the terms $a_1, a_2, \beta_1,
\beta_2, c_1, d_3$ and $d_4$.  Using the transformation (\ref{eq:tr00a}) the
terms $\beta_1$ and $\beta_2$ can be exchanged for $b_1$ and $b_2$,
and the terms $a_2$ and $b_1$ can then be eliminated
using the transformations parameterized by $\sigma_3$ and $\sigma_8$.
This yields
our final action (\ref{eq:details:finalaction}) but without the terms
$e_1$ and $e_2$.  Therefore, for a fully general theory, one can
choose to eliminate higher derivative terms, or stress-energy terms,
but not both.

\subsection{Extension to $N$ scalar fields: Qui-N-tessence}

The preceding analysis can be generalized straightforwardly to the
case of $N$ scalar fields, which we call ``qui-N-tessence'',
an analog of multifield inflation \cite{wands2007,Senatore:2010wk}.
The zeroth order action (\ref{eq:S0}) gets replaced by a general
nonlinear
sigma model:
\begin{align}
S_0 = \int d^4 x \sqrt{-g}  \left[ \frac{\mpl^2}{2} R - \frac{1}{2}
  q_{AB}(\phi^C) \nabla_\nu \phi^A \nabla_\mu \phi^B g^{\mu\nu} -
  U(\phi^C)\right],
\label{eq:S0M}
\end{align}
where $\phi^A = (\phi^1, \ldots, \phi^N)$ are the $N$ scalar fields
and $q_{AB}$ is a metric on the target space.  In the remainder of the
action, functions of $\phi$ are replaced by functions of $\phi^A$.
The first three terms of the second line of Eq.\ (\ref{eq:details:finalaction}) are
replaced by
\begin{align}
&a_{1\, ABCD} \nabla_\mu \phi^A \nabla_\nu \phi^B \nabla_\lambda \phi^C
\nabla_\sigma \phi^D g^{\mu\nu} g^{\lambda\sigma}
+  a_{2\, ABC} \nabla_\mu \nabla_\lambda \phi^A \nabla_\sigma \phi^B
\nabla_\lambda \phi^C g^{\mu\nu} g^{\lambda\sigma}
\nonumber \\
&
+  c_{1\, AB} G^{\mu \nu} \nabla_\mu \phi^A \nabla_\nu \phi^B.
\end{align}
Thus the coefficients $a_1$, $a_2$ and $c_1$ become tensors on the target
space of the indicated orders.
Note that we must use the representation
involving the coefficients $a_{1\,ABCD},a_{2\,ABC}$ and not
$b_{1\,AB},b_{2\,AB}$ (we assume $\alpha_{,A} \ne 0$) since the latter
are less general; the
equivalence between the different representations discussed in Sec.\
\ref{sec:details:final} does not generalize to the $N$ field case.
When $N \ge 4$ one could also add a term
\begin{align}
a_{4\,ABCD} \nabla_\mu \phi^A \nabla_\nu \phi^B \nabla_\lambda \phi^C
\nabla_\sigma \phi^D \epsilon^{\mu\nu\lambda\sigma},
\label{eq:breakparity}
\end{align}
where $a_{4\,ABCD}$ is an arbitrary 4-form on the target space.

\ifx\doejects\undefined
\else
        \eject
\fi
\section{Order of Magnitude Estimates and Domain of Validity}

In the previous sections, we started from the standard quintessence
model with a matter coupling (\ref{eq:S0}), and added arbitrary
corrections involving the scalar field and metric in a derivative
expansion up to four derivatives.  We then exploited the
field-redefinition freedom to eliminate all terms that give
rise to additional degrees of freedom (``higher derivative terms''),
and to reduce the set of operators in the action
to the canonical and unique set given in our final action
(\ref{eq:details:finalaction}).

We now turn to estimating the scaling of the coefficients in the final
action using effective field theory.  We will then use these estimates
to determine the domain of validity of the theory.

\subsection{Derivation of Scaling of Coefficients}
\label{sec:scaling}

We start by recalling the scenario of pseudo Nambu-Goldstone Bosons
(PNGBs)
\cite{1995PhRvL..75.2077F,2003JCAP...07..003A}
discussed in the Introduction that
may give rise to the zeroth order action (\ref{eq:S0}).
Suppose that at some high energy scale $M_*$ we spontaneously break a
set of continuous global symmetries and thereby generate $N$ massless
Goldstone bosons
$\phi^A = (\phi^1, \ldots, \phi^N)$.  The theory then has
$N$ residual continuous symmetries.
If we now suppose that these residual symmetries are explicitly broken
at some much lower energy scale $\Lambda$, then a potential is
generated that scales as
\begin{align}
\Lambda^4 V(\phi^A/M_*),
\end{align}
for some function $V$ which is of order unity.  In particular
the mass of the PNGB fields scale as $\Lambda^2/M_*$ and can be very
light.  For example, in axion models $M_* \sim 10^{12}$ GeV and $\Lambda
\sim \Lambda_{\rm QCD} \sim 100 $ MeV, giving an axion mass of order
$10^{-5}$ eV.

The leading order action for the PNGB fields coupled to gravity at low energies will
be that of a nonlinear sigma model,
\begin{align}
S = \int d^4 x \sqrt{-g}  \left[ \frac{1}{2} \mpl^2R - \frac{1}{2}
  q_{AB}(\phi^C/M_*) \nabla_\nu \phi^A \nabla_\mu \phi^B g^{\mu\nu} -
  \Lambda^4 V(\phi^C/M_*)\right],
\label{eq:S0M1}
\end{align}
where $q_{AB}$ is a metric on the target space
which admits $N$ Killing vectors (the residual symmetries).
In the special case where $q_{AB}$ is flat, these residual symmetries
are shift symmetries $\phi^A \to \phi^A + $ constant.
  We now assume that
these fields drive the cosmic acceleration, and in addition we assume
that the kinetic and potential terms are of the same order, that is,
we assume that slow roll parameters are only modestly small.  It then
follows from the action
(\ref{eq:S0M1}) that the scales of spontaneous and
explicit symmetry breaking
$M_*$ and $\Lambda$ must be of order\footnote{The need to use the
  Hubble scale today in the symmetry breaking scale
  $\Lambda$ is associated with the coincidence problem.}
\begin{align}
M_* \sim \mpl, \ \ \ \ \Lambda \sim \sqrt{H_0 \mpl},
\end{align}
where $H_0$ is the Hubble parameter,
so that the quintessence fields have mass $\sim H_0$ and energy
density $\sim \mpl^2 H_0^2$.
Defining the dimensionless fields $\varphi^A=\phi^A/\mpl$ allows us to
rewrite the action as
\begin{align}
S = \int d^4 x \sqrt{-g}  \left[ \frac{1}{2} \mpl^2R - \frac{1}{2}
 \mpl^2 q_{AB}(\varphi^C) \nabla_\nu \varphi^A \nabla_\mu \varphi^B
 g^{\mu\nu} -   \mpl^2 H_0^2 V(\varphi^C)\right].
\label{eq:S0M2}
\end{align}

Consider now the stability of the theory (\ref{eq:S0M2})
under loop corrections.  The story is exactly the same here as in
inflationary models
\cite{Baumann:2009ni,1995PhRvL..75.2077F} (aside from couplings to
matter, see below).
Computing loop corrections
starting from the action (\ref{eq:S0M2}) does not lead to large
corrections $\delta m \gg H_0$ to the mass of the quintessence fields, because
in the limit where the explicit symmetry breaking scale $\Lambda =
\sqrt{\mpl H_0}$ goes to zero, the theory possesses exact symmetries
which must be respected by the loop corrections.  Hence the loop
corrections to the potential must scale proportionally to $H_0^2
\mpl^2$, as for the original potential.
Thus the smallness of the mass of the quintessence field is
natural in the sense of 't Hooft.
However, this is not the entire story, since the form (\ref{eq:S0M2}) of the low
energy theory imposes non trivial constraints on the physics at high
energies, which must respect the residual symmetries.  Indeed in
general there is no guarantee that there exists a consistent high
energy theory with the low energy limit (\ref{eq:S0M2}).
This question is beyond the scope of this paper: we shall simply
assume that a consistent UV theory can be found.  See Ref.\
\cite{Panda:2010uq} for an example of an attempt to address this issue.

So far in the discussion we have neglected coupling to matter.
If we assume the validity of the weak equivalence principle, the
general leading order coupling of $\phi^C$ to matter will be of the
form of a scalar-tensor theory, given by adding to the action
(\ref{eq:S0M2}) the term
\begin{equation}
S_{\rm m} \left[ e^{\alpha(\phi^c/M_*)} g_{\mu \nu},  \psi_{\rm
    m}\right] = S_{\rm m} \left[ e^{\alpha(\varphi^c)} g_{\mu \nu},  \psi_{\rm
    m}\right],
\end{equation}
for some function $\alpha$.

We now suppose that one or more of the PNGB fields has a mass $\sim M$
which is parametrically larger than $H_0$, and we integrate out
these heavier fields, following the similar analysis of inflationary
models by Burgess, Lee and Trott \cite{Burgess:2009ea}.
Integrating out the heavier fields
gives rise to modifications to the
target space metric and potential for the remaining light fields [that
do not change the scalings
shown in Eq.\ (\ref{eq:S0M2})], and also a set of correction terms to
the leading order action
(\ref{eq:S0M2}).  The leading, tree-level correction terms can be
obtained simply
by solving the classical equations of motion for the heavy fields in
an adiabatic approximation and substituting back into the action.
One finds that the induced correction terms have the
form\footnote{These are the terms involving just the scalar field and
  metric.  One also finds correction terms involving the matter stress
  energy tensor as long as $\alpha^\prime \ne 0$, of the form
  indicated in Table \ref{tab:scalings}.}
\begin{align}
M^2 \mpl^2 \sum_n \frac{c_n}{M^d} {\cal O}_{n},
\label{rule}
\end{align}
where the sum is over operators ${\cal O}_{n}$ involving $d$
derivatives acting on $k$ powers of the dimensionless fields
$\varphi$ and/or $g_{\mu\nu}$, and the coefficients $c_n$ are of
order unity (see Appendix \ref{app:integrateout} for details).  In
other words, each additional derivative is suppressed
by a power of the mass $M$ of the fields that have been integrated out
(which we can think of as a cutoff scale), and the overall prefactor
is such that the normal kinetic terms would be reproduced for the case
$k = d =2$.

\begin{table}
\centering
\begin{tabular}{|c|c|c|}
\hline
Coefficient &Term in Action& Scaling \\
\hline
\hline
$a_1\ \ \ \ $ & $(\nabla \phi)^4$& $\sim 1/(\mpl^2 M^2)$ \\
\hline
$a_2\ \ \ \ $ &$\square \phi (\nabla \phi)^2$ & $\sim 1/(\mpl M^2)$ \\
\hline
$a_3\ \ \dagger \ $ &$(\square \phi)^2$ &  \\
\hline
$b_1\ \ \ \ $ &$T^{\mu\nu} \nabla_\mu \phi \nabla_\nu \phi$ & $\sim
1/(\mpl^2 M^2)$\\
\hline
$b_2\ \ \ \ $ & $T (\nabla \phi)^2$& $\sim 1/(\mpl^2 M^2)$\\
\hline
$b_3\ \ \dagger \ $ &$T \square \phi$ &  \\
\hline
$b_4\ \ \dagger \ $ &$T^{\mu\nu} \nabla_\mu \nabla_\nu \phi$ &  \\
\hline
$b_5\ \ \dagger \ $ &$R^{\mu\nu} T_{\mu\nu}$ & \\
\hline
$b_6\ \ \dagger \ $ &$R T$ &  \\
\hline
$b_7\ \ \ \ $ &$T$ &   \\
\hline
$c_1\ \ \ \ $ &$G^{\mu\nu} \nabla_\mu \phi \nabla_\nu \phi$ & $\sim
1/M^2$ \\
\hline
$c_2\ \ \dagger \ $ &$R (\nabla \phi)^2$ &   \\
\hline
$c_3\ \ \dagger \ $ &$R \square \phi$ & \\
\hline
$d_1\ \ \dagger \ $ &$R^2$ &   \\
\hline
$d_2\ \ \dagger \ $ &$R^{\mu\nu} R_{\mu\nu}$ &  \\
\hline
$d_3\ \ \ \ $ &Gauss-Bonnet &  $\sim \mpl^2/M^2$ \\
\hline
$d_4\ \ \ \ $ &Chern-Simons & $\sim \mpl^2/M^2$ \\
\hline
$e_1\ \ \ \ $ &$T^{\mu\nu} T_{\mu\nu}$ & $\sim 1/(\mpl^2 M^2)$ \\
\hline
$e_2\ \ \ \ $ &$T^2$ & $\sim 1/(\mpl^2 M^2)$ \\
%\hline
%Coefficient &Term in ${\bar g}_{\mu\nu}$& & \\
%\hline
%$\beta_1\ \ \ \ $ &$\nabla_\mu \phi \nabla_\nu \phi$ & $\sim 1/M^4$
%&$\lesssim 1/(\mpl^2 M^2)$ \\
%\hline
%$\beta_2\ \ \ \ $ &$(\nabla \phi)^2 g_{\mu\nu}$ & $\sim 1/M^4$ &
%$\lesssim 1/(\mpl^2 M^2)$\\
%\hline
%$\beta_3\ \ \ \ $ &$\square \phi g_{\mu\nu}$ & $\sim \mpl/M^4$ & \\
%\hline
%$\beta_4\ \ \ \ $ &$\nabla_\mu \nabla_\nu \phi$ & $\sim \mpl/M^4$ & \\
%\hline
%$\beta_5\ \ \ \ $ &$R_{\mu\nu}$ & $\sim \mpl^2/M^4$ & \\
%\hline
%$\beta_6\ \ \ \ $ &$R g_{\mu\nu}$ & $\sim \mpl^2/M^4$ & \\
\hline

\end{tabular}
\caption{This table gives the scalings of the various coefficients.
The first column lists the coefficients, and the second column lists the
corresponding terms in the
  action (\protect{\ref{eq:S1}}).
Daggers in the first column indicate higher derivative terms.
The third column gives our
estimate of the scale of the coefficients, under the
assumptions discussed in the
text, for those coefficients that are nonzero in our final action
(\protect{\ref{eq:details:finalaction}}), or
in versions of that action obtained using the
field redefinitions (\protect{\ref{eq:ch3}}) or
(\protect{\ref{eq:ch8}}).
The quantity $M$ is the mass of the lightest field that is integrated
out to produce our final action.
In all cases, these scales for the coefficients correspond
to fractional corrections to the leading order dynamics of order
$\sim H_0^2/M^2$.}
\label{tab:scalings}
\end{table}

Note that the rule (\ref{rule}) for how the coefficients of additional
corrections to the action depend on the cutoff scale $M$ differs from
the usual rule of effective field theory, where an operator of
dimension $D+4$ has a coefficient $\sim M^{-D}$.
The rule (\ref{rule}) instead gives a coefficient $\sim M^{-(d-2)}
\mpl^{-(k-2)}$, where $d$ is the number of derivatives in the operator
and $k$ is the number of powers of (canonically normalized) fields,
related to $D$ by $D = d+k-4$.
The difference between the two rules
arises from the fact that we are making nontrivial assumptions about
the physics above the scale $M$,
specifically that it is described by an action of the PNGB form
(\ref{eq:S0M2}) \footnote{More general interactions which are not of
the form (\ref{eq:S0M2}) can modify the scaling rule (\ref{rule}), even
if they respect the residual (shift) symmetries.
For example consider a scalar field $\psi$ of mass $m$ which couples to $\phi$ via a term $\psi (\nabla \phi)^2/m_*$ for some mass scale $m_*$.
Integrating out this field gives a correction to the $\phi$ action $\sim (\nabla \phi)^4/(m^2 m_*^2)$ (see Appendix \ref{app:integrateout}).
To keep such terms from invalidating the scaling rule we need to
assume that $m m_* \gtrsim M \mpl$, i.e. that any such fields are
either sufficiently massive or sufficiently weakly coupled to the PNGB fields.}.
If we were to allow arbitrary physics at energies
above
the scale $M$, then the coefficients would scale according to the
standard rule.

We now specialize to the case of a single light field.  The correction
terms (\ref{rule}) have the form of a double power
series, in number of derivatives and in powers of the fields.
If we fix the number of derivatives and associated index structure, we
can sum over all operators that differ only by powers of $\varphi$
to obtain operators with prefactors that are functions of $\varphi$,
$
f(\varphi) = \sum c_k \varphi^k
$
with coefficients of order unity.
We now write out all the resulting
terms to leading order in $1/M^2$, imposing general covariance.  The
result is the theory
(\ref{eq:details:finalaction})
discussed in the last section\footnote{The parity-violating
Chern-Simons term is not generated in this way, since the fields we
are integrating out do not violate parity.  To obtain the
Chern-Simons term with the scaling indicated in Table
\protect{\ref{tab:scalings}} would require integrating out some
parity violating fields at the scale $M$ which approximately respect
the residual (shift) symmetries.}, but with additional information
about the coefficients $a_1$, $b_1$ etc.  Specifically we find that
\begin{align}
a_1(\phi) = \frac{1}{\mpl^2 M^2} {\hat a}_1(\phi/\mpl),
\end{align}
where the function ${\hat a}_1$ is of order unity, i.e., the
coefficients in its Taylor expansion are
independent of $\mpl$ and $M$.  The corresponding prefactors or
overall scaling for the other coefficients are listed in
Table
\ref{tab:scalings}.

Finally, we note that, as is well known,
Solar System tests of general relativity strongly constrain
the coupling of $\phi$ to the matter sector\footnote{Strictly
  speaking, Solar System tests lie outside the domain of validity of
  our effective field theory  unless $M^{-1} \lesssim 1 \, {\rm
    A.U.}$, which is very small compared to $H_0^{-1}$; see Sec.\
  \protect{\ref{sec:validity}} above.}.
If we define the dimensionless parameter $\lambda = \mpl |\alpha^\prime(\phi_0)|$, where $\phi_0$ is the present day cosmological
background value of $\phi$, then the Solar System constraint
is\footnote{This constraint can be evaded in models where
  nonlinear effects in $\phi$ are important in the Solar System, such as
  Chameleon \protect{\cite{Khoury2004}} and Galileon models
  \protect{\cite{Nicolis2009,Deffayet:2009wt,Deffayet:2010zh}}.}
$\lambda \lesssim 10^{-2}$ \cite{Will:2005va}.
In addition the coupling of the scalar to the visible sector will
generically give rise to large corrections to the quintessence
potential via loop corrections \cite{Brax:1999yv,Doran:2002qd,Pietroni:2005pv,Garny:2006wc,Arbey:2007vu}.
For a fermion
of mass $m_f$, the
correction $\delta m$ to the mass of the quintessence field will be of
order
\begin{align}
\frac{\delta m}{H_0} \sim \lambda \left( \frac{m_f^2}{H_0 \mpl} \right).
\end{align}
If $\lambda \sim 1$ and
$m_f \gg \sqrt{\mpl H_0} \sim 10^{-3}$ eV,
then $\delta m \gg H_0$, which is inconsistent if the quintessence
field is to drive cosmic acceleration.
This is a well known naturalness problem for matter couplings in
quintessence models, and
it motivates setting\footnote{More precisely the condition is
  $\alpha^\prime =0$, i.e., $\alpha = $ constant, but
  the constant can be absorbed by a rescaling of all the dimensionful
  parameters in the matter action.} $\alpha = 0$.

\subsection{Domain of Validity of the Effective Field Theory}
\label{sec:validity}

We now estimate the domain of validity for the theory
(\ref{eq:details:finalaction}) with
the scalings given by Table \ref{tab:scalings},
by requiring that the terms with higher derivatives be small compared
to terms with fewer derivatives.
If $E$ is the energy involved in a given process, or equivalently
$E^{-1}$ is the corresponding time-scale or length-scale, then successive
terms in the derivative expansion are suppressed by the ratio $E/M$,
which yields the standard condition
\begin{align}
E \ll M
\label{eq:con1}
\end{align}
for the domain of validity.
As discussed in the Introduction,
$M$ must be somewhat larger than $H_0$ in order to describe the background
cosmology and observable perturbation modes.  However if $M$ is
significantly larger than $H_0$ then the corrections due to the higher
order terms in Eq.\ (\ref{eq:details:finalaction}) become negligible,
and the theory reduces to a standard quintessence model with some
matter coupling.  Therefore, the interesting regime is when
$M$ is perhaps just one or two orders of magnitude larger than $H_0$,
as indicated in Fig.\ \ref{fig:scales0}.
In particular, when the scale $M$ is in this interesting regime, the
theory is unable to describe gravitational effects in the Solar
System and binary pulsars, which is a shortcoming of the effective
field theory approach used here.

Consider now the background cosmological solution.  The theory
(\ref{eq:start}) to zeroth order in $\epsilon$ (or equivalently $1/M^2$)
has the equations of motion
\bes
\begin{align}
\label{eq:eom0a}
  \mpl^2 G_{\alpha\beta} = {}& \nabla_\alpha \phi \nabla_\beta \phi -
  \frac{1}{2} (\nabla \phi)^2 g_{\alpha\beta} - U(\phi)
  g_{\alpha\beta} + e^{2 \alpha(\phi)} T_{\alpha\beta},
\\
  \square \phi = {}& U^\prime (\phi) - \frac{1}{2} \alpha^\prime e^{2
    \alpha} T.
\label{eq:eom0b}
\end{align}
\ees
For each of these two equations we assume that all of the terms
are of the same order. For the matter terms this is
this is a reasonable approximation, since $\Omega_\Lambda \sim 0.7$
and $\Omega_{\mathrm{matter}} \sim 0.3$. If
the scalar potential term dominates over the kinetic term,
then the following estimates need to be modified by including factors
of slow roll parameters; we ignore these factors here since we expect
them to be only modestly small.
Similarly, our estimates assume that $\mpl \alpha^\prime$ is of order unity;
some changes would be required if this quantity were very small.
From these assumptions, and ignoring $O(1)$ functions of the scalar
field, we have
\begin{align}
  & \mpl^2 R \sim  (\nabla \phi)^2 \sim U \sim T \sim
 \mpl \square \phi \sim \mpl U^\prime(\phi) \sim
H_0^2 \mpl^2.
\end{align}
Inserting these estimates into the action
(\ref{eq:details:finalaction}) and using the
scalings given in Table \ref{tab:scalings},
we find that for each of the correction terms in the action,
the fractional corrections to the leading
order cosmological dynamics scale as $H_0^2/M^2$.
The corrections therefore are of
order unity at $M \sim H_0$, as we would expect, since at this scale
the heavy fields which we have integrated out have the same mass scale
as the light fields, and would be expected to give rise to $O(1)$
corrections to the dynamics.  This gives a useful consistency check of
the calculations underlying Table \ref{tab:scalings} discussed in
the previous subsection.

In addition to the standard constraint (\ref{eq:con1}), there are
other constraints on the domain of validity which we now discuss.
We focus attention on cosmological perturbations, for which
$\phi(t,{\bf x}) = \phi_0(t) + \delta \phi(t,{\bf x})$, and consider
the conditions under which the dynamics of the perturbation $\delta
\phi$ can be described by the effective theory.
Consider localized wavepacket modes $\delta \phi$, where the size of
the wavepacket is of the same order as the wavelength, both $\sim
E^{-1}$.
For such modes we can characterize perturbations in terms of two
parameters, the energy $E$ and the number of quanta or mode occupation
number $N$.  The total energy of the wavepacket will be of order
$
N E \sim \int d^3 x ({\bf \nabla} \delta \phi)^2 \sim E^{-3} (E \delta
\phi)^2
$
which gives the estimate
\begin{align}
\delta \phi \sim \sqrt{N} E.
\end{align}
The fractional density perturbation due to the wavepacket is of
order
\begin{align}
\frac{\delta\rho}{\rho} \sim \frac{({\bf \nabla} \delta \phi)^2 }{
  H_0^2\mpl^2} \sim
\frac{N E^4}{H_0^2 \mpl^2}.
\label{eq:fdp}
\end{align}
We now demand that the term $a_1 (\nabla \delta \phi)^4$ in the
action\footnote{Here we envisage computing an action for the
  perturbations by expanding the action
  (\protect{\ref{eq:details:finalaction}}) around the background
  cosmological solution, as in Ref.\ \protect{\cite{Burgess:2009ea}}.}
be small compared to the leading order term $(\nabla \delta \phi)^2$.
Using the scaling $a_1 \sim 1/(\mpl^2 M^2)$ from Table
\ref{tab:scalings} and combining with the estimate (\ref{eq:fdp}) of the
fractional density perturbation then gives the constraint\footnote{In the previous subsection we showed that $a_1(\phi) = {\hat a}_1(\phi/\mpl) / (M^2 \mpl^2)$,
where ${\hat a}_1$ is function for which all the Taylor expansion coefficients
are of order unity.  It follows that ${\hat a}_1 \sim 1$ for $\phi \sim \mpl$.
However the estimate (\protect{\ref{eq:con2a}}) requires the stronger
assumption ${\hat a}_1 \lesssim 1$ for $\phi \gg \mpl$ which need not
be valid.  If we instead assume that ${\hat a}_1 \sim
(\phi/\mpl)^\alpha$ for $\phi \gg \mpl$ then the constraint
(\protect{\ref{eq:con2b}}) gets replaced by $N (E/M)^\gamma \ll
\mpl^2/M^2$, where $\gamma = 2(4+\alpha)/(2+\alpha)$.  This modifies
the boundary of the domain of validity of the effective field theory
shown in Fig.\ \protect{\ref{fig:scales1}} by changing the slope of
the tilted portion of the boundary.  In the limit $\alpha \to \infty$
this portion of the boundary approaches the green curve $\delta
\varphi \sim \mpl$.}
\begin{align}
\frac{\delta \rho}{\rho} \ll \frac{M^2}{H_0^2}.
\label{eq:con2a}
\end{align}
Thus, the theory can describe perturbations in the nonlinear regime,
but the perturbations can only be modestly nonlinear if $M$ is fairly
close to $H_0$.  In terms of the parameters $E$ and $N$ the constraint
(\ref{eq:con2a}) is
\begin{align}
N E^4 \ll M^2 \mpl^2.
\label{eq:con2b}
\end{align}
This gives a nontrivial constraint on the domain of validity of the
theory in the regime $E \lesssim M$.
The two dimensional parameter space $(E,N)$ is illustrated in Fig.\
\ref{fig:scales1}, which shows the constraints (\ref{eq:con1}) and (\ref{eq:con2a}),
the curves $\delta \rho/\rho \sim 1$ and $\delta \rho \sim M^2/H_0^2$,
as well as the curve where $\delta \phi \sim \mpl$.

Another potential constraint on the domain of the validity of the
theory (\ref{eq:details:finalaction}) with the scalings given by Table
\ref{tab:scalings} is that the theory should be weakly coupled,
i.e. the effects of loop corrections should be small.  Using the power
counting methods of Ref.\ \cite{Burgess:2009ea} one can show that
this is indeed true within the domain $H_0 \lesssim E \ll M$ of interest.
Strong coupling can arise due to tri-linear couplings, as discussed in
Sec. 2.2 of Ref.\ \cite{Burgess:2009ea}, but this only occurs for
energies far below the Hubble scale $H_0$, and so is not relevant to
cosmological applications of the theory.

We note that there are several well known theories of cosmic
acceleration that are not encompassed by our effective field theory. The form of our expansion requires that the
dominant contribution to cosmic acceleration be the
leading order scalar terms and not the higher order terms, and so theories in which other mechanisms provide the acceleration cannot be described in our formalism.
One example is provided by K-essence models in which terms in the
action like $(\nabla \phi)^4$, $(\nabla \phi)^6$ \ldots are all
equally important.  In particular this is true for ghost condensate
models \cite{Arkani-Hamed2004}.  Also there are many cosmic
acceleration models that exploit the Vainshtein effect
\cite{Vainshtein1972, Luty2003,Alberto2004} to evade Solar
System constraints on light fields with gravitational-strength couplings.
The Vainshtein effect relies on nonlinear derivative terms in the
scalar field action.  Although our class of theories includes models that
demonstrate the Vainshtein mechanism, the mechanism only operates
outside the domain of validity of our approach, as we require the nonlinear
derivative terms to be small.
The chameleon mechanism \cite{Khoury2004,Khoury2004a}, on
the other hand, does not require nonlinearities in the derivatives of
the scalar field, and thus may be analyzed in our formalism, although
the regime in which a screening mechanism would be required to evade
fifth force experiments and solar system constraints will be in the
regime of validity of our analysis only for large enough values of
the cutoff $M$.

\ifx\doejects\undefined
\else
        \eject
\fi
\section{Discussion and Conclusions}
\label{sec:conclusion}

In this paper, we have investigated effective field theory models of
cosmic acceleration involving a metric and a single scalar field.
The set of theories we considered consists of a standard quintessence
model with matter coupling, together with a general covariant
derivative expansion, truncated at four derivatives.  We showed that
this class of theories can be obtained from a PNGB scenario, where one
of the PNGB fields is lighter than all the others, and the heavier
fields are integrated out.  We showed that in constructing this class
of theories, including higher derivative terms in the action, as
suggested by Weinberg \cite{Weinberg2008}, does not give any
increased generality.  We also showed that complete generality
requires one to include terms in the action that depend on the
stress-energy tensor of the matter fields.

We now turn to a discussion of
some of the advantages and shortcomings
of the approach adopted here to describe models of dark energy.  Some
of the shortcomings are:

\begin{itemize}

\item By construction, our approach excludes theories where nonlinear
  kinetic terms in the action give an order unity contribution to the
  dynamics, such as K-essence, ghost condensates etc., since such theories
  do not arise from the PNGB construction used in this paper.
  On the other hand, such theories are less natural than the class of theories
  considered here, from the point of view of loop corrections: they
  require very nontrivial physics at the scale $\sim H_0$, instead of
  at the scale $\sim \sqrt{H_0 \mpl}$ required in the PNGB approach.
  The most general class of theories of this kind is that of Horndeski
  \cite{Horndeski1974}, which contains four free functions of $\phi$
  and $(\nabla \phi)^2$ \cite{Skordis2011}, and which is the most
  general class of theories of a metric and a scalar field for which
  the equations of motion are second order.
  As discussed in the Introduction, these theories are included
  in the alternative, background-dependent approach to effective
  field theories of quintessence of Creminelli et al. \cite{Creminelli2009}.

\item Our class of theories will be observationally distinguishable
  from vanilla quintessence theories only if the cutoff $M$ is near
  the Hubble scale $H_0$.  In this regime, our framework cannot be
  used to analyze Solar System tests of general relativity, since they are
  outside the domain of validity of the effective field theory.
  Also, when the background cosmology is evolved backwards in time it
  passes outside the domain of validity at fairly low redshifts. (This is
  not a serious disadvantage since dark energy dominates only at low
  redshifts.)

\item We have restricted attention to theories with a metric and a
single scalar field, with the only symmetry being general covariance.  Thus,
our analysis does not include models with several scalar fields,
vector fields etc.  In addition, our analysis excludes an interesting
class of models
that one obtains by imposing that the action be invariant under $\phi
\to f(\phi)$, where $f$ is any monotonic function, as such a symmetry cannot be realized with our derivative expansion.  This class of
models includes Horava-Lifshitz gravity and has the same number of physical degrees of freedom as general relativity \cite{Skordis2011,Blas:2010hb}.
It would be interesting to explore the most general dark energy models
of this kind.

\end{itemize}

Some of the advantages of the approach used here are:

\begin{itemize}

\item Our class of theories is generic within the PNGB construction,
  which itself is a well motivated way to obtain the ultralight fields
  needed for cosmic acceleration.  The theories are fairly simple and
  it should be straightforward to confront them with observational data.

\item Our class of theories allow for a unified treatment of the
  cosmological background and perturbations, unlike the
  background-dependent approach of Ref.\ \cite{Creminelli2009}.

\end{itemize}

Finally, we list some possible directions in which the approach used
here could be extended:

\begin{itemize}

\item It would be interesting to compute the relation between the nine
  free functions used in our theories to the free functions of
the post-Friedmannian approach to parameterizing dark energy models \cite{Skordis2011}.

\item It would be interesting to explore the phenomenology of the
  various higher order terms in our action, for the cosmological
  background evolution and perturbations.  Many of the terms have
  already been explored in detail, see for example Refs.\
  \cite{Deffayet:2010qz,Charmousis2011}.

\item Either by using the post-Friedmannian approach, or more
  directly, it would be useful to compute the current observational
  constraints on the free functions in the action.

\item An interesting open question is the extent to which our final
  action is generic.  That is, is there a class of theories more
  general than nonlinear sigma model PNGB theories for which our
  action is obtained by integrating out some of the fields?

\end{itemize}

\subsubsection*{Acknowledgments}

We thank Justin Vines, Liam McAllister, Guido D'Amico and Scott Watson
for helpful discussions. This research was supported in part by
NSF grants PHY-0757735 and PHY-0968820, by NASA grants
NNX08AH27G and NNX11AI95G, and by
the John and David Boochever Prize Fellowship in Fundamental
Theoretical Physics to JB at Cornell.

\bibliography{darkenergy}
\bibliographystyle{JHEP}

\newpage
\clearpage

\appendix
\section{Notation and Conventions}
\label{app:notation}

We use natural units with $c = \hbar = 1$, and define the
the reduced Planck mass via $\mpl^2 = 1/8\pi G$.
The Einstein and Jordan frame metrics are $g_{\mu\nu}$ and ${\bar
  g}_{\mu\nu}$ respectively, and the corresponding derivative
operators are $\nabla_\mu$ and ${\bar \nabla}_\mu$.  We use the usual
abbreviations $(\nabla \phi)^2 = g^{\mu\nu} \nabla_\mu \phi \nabla_\nu
\psi$ and $\square \phi = \nabla_\mu \nabla^\mu \phi$.
Primes denote derivatives with respect to the scalar field $\phi$, as
in $U^\prime (\phi)$.  We use the $(-,+,+,+)$ metric signature and the
sign conventions $(+,+,+)$ in the notation of Ref.\
\cite{1973grav.book.....M}.   Finally we take $\epsilon^{\mu \nu
  \lambda \rho}$ to be the antisymmetric tensor with $\epsilon^{0123}
= 1/\sqrt{-g}$.

We define the (Jordan-frame) stress-energy tensor $T_{\mu}^{\ \,  \nu}$ in the usual way in terms of the
Jordan-frame metric ${\bar g}_{\mu\nu}$ that appears in the matter
action $S_{\rm m}$:
\begin{align}
S_{\rm m}[{\bar g}_{\mu\nu} + \delta {\bar g}_{\mu\nu},\psi_{\rm m}]
-S_{\rm m}[{\bar g}_{\mu\nu},\psi_{\rm m}]
=
\frac{1}{2} \int d^4 x \sqrt{- {\bar g}} T_{\mu}^{\ \, \nu} {\bar
  g}^{\mu\lambda} \delta {\bar g}_{\lambda\nu} + O(\delta {\bar g}^2).
\label{eq:Tabdef}
\end{align}
We then define $T = T_{\mu}^{\ \, \mu}$, and define the quantities $T_{\mu\nu}$ and $T^{\mu\nu}$ by raising
and lowering indices with the Einstein-frame metric $g_{\mu\nu}$,
which is related to ${\bar g}_{\mu\nu}$ via Eq.\
(\ref{eq:JordanMetric}).  To zeroth order in $\epsilon$ this stress
energy tensor obeys the conservation law
\begin{align}
e^{-2 \alpha} \nabla_\lambda (e^{2 \alpha} T^{\lambda\sigma} ) = \frac{1
 }{2} \alpha' T \nabla^\sigma \phi + O(\epsilon).
\end{align}

\ifx\doejects\undefined
\else
        \eject
\fi
\section{The Weak Equivalence Principle}
\label{app:WEP}

In this Appendix, we show that including terms in the action that
depend explicitly on the matter stress energy tensor, as in Eq.\
(\ref{eq:start}) above, generically gives rise to violations of the weak equivalence
principle.  However, we also show that our specific model
(\ref{eq:start}) does not, to linear order in $\epsilon$.
Since the parameter $\epsilon$ essentially counts the number of
derivatives in our derivative expansion, it follows the weak equivalence
principle is satisfied for our derivative expansion up to four
derivatives.

\subsection{Generic Violations of Weak Equivalence Principle when Stress-Energy Terms are Present in Action}

Consider first an action principle of the general form
\begin{align}
S[g_{\alpha\beta},\phi,\psi_{\rm m}] = S_{\rm g}[g_{\alpha\beta},\phi]
+ S_{\rm  m}[{\bar g}_{\alpha\beta},\psi_{\rm m}].
\label{eq:WEP}
\end{align}
Here the first term is a gravitational action, depending only on the
metric $g_{\alpha\beta}$ and the scalar field $\phi$, and the second
term is the matter action, in which all the matter fields $\psi_{\rm
  m}$ couple only to the Jordan metric ${\bar g}_{\alpha\beta}$ (some
function of $g_{\alpha\beta}$ and $\phi$), and not to
$g_{\alpha\beta}$ and $\phi$ individually.
By definition, any theory of this form obeys the weak equivalence
principle.  What this means is as follows.  We define weakly
self-gravitating bodies to be bodies for which we can neglect the
perturbations they cause to $g_{\alpha\beta}$ and $\phi$.  From the
form of the action (\ref{eq:WEP}), it follows that all weakly
self-gravitating bodies will fall on geodesics of the metric ${\bar
  g}_{\alpha\beta}$, and hence will all fall on the same geodesics.

The action principle (\ref{eq:start}) we use in this paper is not of
the general form (\ref{eq:WEP}), because of the explicit appearance of
terms involving the stress energy tensor in the gravitational action.
Therefore one expects violation of the weak equivalence principle to
arise.  We now verify explicitly that this does occur in a specific
example.
We choose the following special case of the action (\ref{eq:start}),
where the only perturbative term included is the term proportional to
the trace of the stress energy tensor:
\begin{align}
S = \int d^4 x \sqrt{-g} \left[ \frac{1}{2} \mpl^2 R - \frac{1}{2}
  (\nabla \phi)^2 - U(\phi) + \epsilon f(\phi) T \right] + S_{\rm
  m}[{\bar g}_{\alpha\beta},\psi_{\rm m}].
\end{align}
We choose the matter field $\psi_{\rm m}$ to be a scalar field $\psi$ with
action
\begin{align}
S_{\rm m} = - \int d^4 x \sqrt{- {\bar g}}  \left[ \frac{1}{2} ({\bar
    \nabla} \psi)^2 + V(\psi) \right],
\end{align}
and we specialize the relation (\ref{eq:JordanMetric}) between the two metrics to be
the conformal transformation ${\bar g}_{\alpha\beta} =
e^{\alpha(\phi)} g_{\alpha\beta}$.  This gives $T = - e^{-\alpha}
(\nabla \psi)^2 - 4 V$ and the total action is therefore
\begin{align}
S = \int d^4 x \sqrt{-g} \left[ \frac{1}{2} \mpl^2 R - \frac{1}{2}
  (\nabla \phi)^2 - U(\phi) - \frac{1}{2} (e^\alpha + 2 \epsilon
  e^{-\alpha} f) (\nabla \psi)^2 - (e^{2 \alpha} + 4 \epsilon f) V(\psi)\right].
\end{align}
The kinetic term for $\psi$ can be written as $\int d^4 x \sqrt{-{\hat
    g}} ({\hat \nabla}\psi)^2$ where ${\hat g}_{\alpha\beta} =
(e^\alpha + 2 \epsilon e^{-\alpha} f) g_{\alpha\beta}$, and the potential term
can be written as $\int d^4 x \sqrt{- {\tilde g}} V(\psi)$, where
${\tilde g}_{\alpha\beta} = \sqrt{e^{2 \alpha} + 4 \epsilon f} g_{\alpha\beta}$.
Therefore, objects whose stress energy is composed of different
combinations of the kinetic term and the potential term will fall on
different combinations of the metrics ${\hat g}_{\alpha\beta}$ and
${\tilde g}_{\alpha\beta}$, violating the weak equivalence principle.

\subsection{Validity of Weak Equivalence Principle to Linear Order}

In the above analysis, we note that the metrics ${\hat g}_{\alpha\beta}$
and ${\tilde
  g}_{\alpha\beta}$ coincide to linear order in $\epsilon$, so there
is no violation to this order.  We now show
that, similarly, none of the stress-energy-dependent terms
included in Eq.\ (\ref{eq:start}) violate the weak equivalence
principle, to linear order in $\epsilon$.

The key idea of the proof is to use the transformation laws derived in
Sec.\ \ref{sec:techs}
above to rewrite the theory in the general form (\ref{eq:WEP}), which we
know satisfies the weak equivalence principle.
All of the terms in the action given by Eqs.\ (\ref{eq:start})  --
(\ref{eq:JordanMetric}) are of this form,
except for the terms parameterized by the coefficients $b_1, \ldots,
b_7$, $e_1$ and $e_2$.
However, as we now show, we can use transformations to eliminate these
terms in favor
of the remaining terms which manifestly satisfy the principle.

Consider first the terms in the action (\ref{eq:S1}) which depend linearly on the
stress-energy tensor.  We can eliminate the terms parameterized by
$b_1, \ldots, b_6$ using
the transformation (\ref{eq:tr00}) with ${\tilde \beta}_i = -2 e^{-2
  \alpha} b_i$ for $ 1 \le i \le 6$.
This generates contributions to the
the terms parameterized by $\beta_1, \ldots , \beta_6$ in the
definition (\ref{eq:JordanMetric}) of the Jordan metric.
Similarly, by using the transformation (\ref{eq:tr01}) with ${\tilde
  \alpha} = -2 e^{-2 \alpha} b_7$, we can eliminate the term
parameterized by $b_7$ in favor of an $O(\epsilon)$ correction to the
function $\alpha$ in Eq.\ (\ref{eq:JordanMetric}).

We now turn to the terms in the
action (\ref{eq:S1}) which depend quadratically on the stress-energy
tensor, namely the terms parameterized by $e_1$ and $e_2$.
For $e_1$ we use the transformation
(\ref{eq:ch11}) with $\sigma_{11}
= - e^{-2 \alpha} e_1$, and
for $e_2$ we use the transformation
(\ref{eq:ch10}) with $\sigma_{10}
= - e^{-2 \alpha} e_2$.
These transformations generates new contributions to the linear
stress-energy terms
parameterized by $b_1$, $b_2$, $b_5$, $b_6$ and $b_7$ (see Table \ref{tab:transformations}), but we have
already shown
that all of those terms satisfy the weak equivalence principle.

To summarize, we have shown that our model (\ref{eq:start}) satisfies
the weak equivalence principle despite the explicit appearance of stress energy
terms in the action.  Of course, there can be violations of the strong
equivalence principle in models of this kind, which can even be of
order unity \cite{Stubbs2009}.  In addition, the weak equivalence
principle will generically be violated by quantum loop corrections,
although this is a small effect \cite{Armendariz-Picon2011}.

\subsection{Potential Ambiguity in Definition of Weak Equivalence Principle}

We next discuss a potential ambiguity that arises in the definition
of the weak equivalence principle.  In the definition one restricts
attention to bodies whose gravitational fields, as measured by the
perturbations they produce to the metric $g_{\mu\nu}$ and scalar field
$\phi$, can be neglected.  However, consider for example the field
redefinition (\ref{eq:ch10}), where the metric transforms according to
\begin{align}
g_{\alpha\beta} = {\hat g}_{\alpha\beta}
+ 2 \epsilon \sigma_{10} T {\hat g}_{\alpha\beta}.
\label{eq:ss}
\end{align}
It is possible for the perturbation $\delta {\hat g}_{\alpha\beta}$
generated by the body to be negligible, but the perturbation $\delta
g_{\alpha\beta}$ to be non-negligible, because of the appearance of
the stress-energy term in Eq.\ (\ref{eq:ss}).  If this occurs then the
weak equivalence principle could be valid for one choice of variables,
but not valid for the other choice.

To assess this ambiguity, we now make some order of magnitude
estimates.  Consider a body of mass $\sim M_b$ and size $\sim R$. Then
in general relativity the size of the metric perturbation due to the
body is of order $\delta {\hat g}_{\alpha\beta} \sim M_b/(\mpl^2 R)$.
Suppose now that $\sigma_{10} \sim 1/(\mpl^2 M^2)$, as indicated by Eq.\
(\ref{eq:tr10}) and Table \ref{tab:scalings}.
Then the contribution to the metric perturbation $\delta
g_{\alpha\beta}$ from the second term in Eq.\ (\ref{eq:ss}) will be of
order $M_b / (R^3 \mpl^2 M^2)$, which will be much larger than $\delta
{\hat g}_{\alpha\beta}$ whenever $R \ll M^{-1}$.
Therefore the ambiguity could in principle arise.

However, in the models considered in this paper the ambiguity does not
occur.  This is because the condition $R \ll M^{-1}$ is excluded by
the condition (\ref{eq:con1}) for the validity of the
effective field theory.

\ifx\doejects\undefined
\else
        \eject
\fi
\section{Equivalence Between Field Redefinitions, Integrating Out New Degrees of Freedom, and Reduction of Order}
\label{app:backsubs}

The action (\ref{eq:start}) we start with in the body of the paper
contains several higher derivative terms, that is, terms which gives
contributions to the equations of motion which involve third-order and
fourth-order time derivatives of
the fields.  As discussed in the
Introduction, the theory with these higher derivative terms contains
additional degrees of freedom compared to our zeroth order action
(\ref{eq:S0}), which contains a single graviton and scalar.
In this paper our goal is to describe a general class of theories
containing just one tensor and one scalar degree of freedom, so we
wish to exclude these additional degrees of freedom\footnote{Higher
derivative terms are also generically associated with instabilities
\cite{Woodard2007}, although this can be evaded in special cases, for example
$R^2$ terms.}.

Therefore, as discussed in the Introduction, we define the theory we
wish to consider, associated with
our action (\ref{eq:start}), to be that obtained from the following
series of steps:

\begin{enumerate}

\item Vary the action to obtain the equations of motion, which will
  contain third-order and fourth-order derivative terms which are
  proportional to $\epsilon$.

\item Perform a {\it reduction of order} procedure on the equations of
  motion \cite{1971ctf..book.....L,Parker:1993dk,Flanagan:1996gw}.
  That is, substitute the zeroth-order in $\epsilon$ equations of motion
  into the higher derivative terms in order to obtain equations that
  contain only second-order and lower order time derivatives, which are
  equivalent to the original equations up to correction terms of
  $O(\epsilon^2)$ which we neglect.

\item Optionally, one can then derive the action principle that gives
  the reduced-order equations of motion.

\end{enumerate}
In this Appendix, we show that this procedure is equivalent to the
computational procedure we use in the body of the paper, in which we
apply perturbative field redefinitions directly to the action in order
to obtain an action with no higher derivative terms.
We also show that it is equivalent to integrating
out at tree level the extra degrees of freedom that are associated
with the higher derivative terms.

We note that the analyses of general quintessence models by
Weinberg \cite{Weinberg2008} and Park {\it et al.} \cite{Watson2010}
used a different method of eliminating higher derivative terms.
They performed a reduction of order procedure directly at the level
of the action, that is, they substituted the zeroth-order equations of
motion directly into the higher derivative terms in the action, to
obtain an action with no higher derivative terms.
We will show that this method is not in general
correct; it does not agree
with the theory obtained by applying the reduction of order method to
the equations of motion\footnote{The reason is that substituting the
  zeroth order equations of motion into the action gives an action
  which is correct off-shell to $O(\epsilon^0)$ and on-shell to
  $O(\epsilon)$, but it needs to be valid off-shell to
  $O(\epsilon)$.}.  However, it differs from the correct result
only by field redefinitions (that do not involve higher derivatives),
and so for the purpose of attempting to
classify general theories of quintessence, Weinberg's method is
adequate.

\subsection{Reduction of Order Method}

We start by considering the case of just a scalar field; a more
general argument valid for scalar and tensor fields will be given below.
Consider a general action of the form
\begin{align}
  S = \int d^4 x \sqrt{-g} \left\{ - \frac{1}{2} (\nabla \phi)^2 -
    U(\phi) + \epsilon F[\phi, (\nabla \phi)^2, \square \phi ]  \right\},
\label{eq:hd}
\end{align}
where $F$ is an arbitrary function.  We introduce the notation $K =
(\nabla \phi)^2$ and $L = \square \phi$.  We first show that applying
the reduction of order procedure to the equations of motion (steps 1
-- 3 above) give rise to a theory of the form (\ref{eq:hd}) but with
$F(\phi,K,L)$
replaced by another function ${\hat F}(\phi,K,L)$, given by
\begin{align}
{\hat F}(\phi,K,L) = F[\phi,K,U'(\phi)] + [L - U'(\phi)]
F_{,L}[\phi,K,U'(\phi)].
\label{eq:hatF}
\end{align}

To see this, we vary the action (\ref{eq:hd}) to obtain the equation of
motion
\begin{align}
\square \phi - U'(\phi) + \epsilon F_{,\phi} - 2 \epsilon
\nabla_\alpha (F_{,K} \nabla^\alpha \phi) + \epsilon \square F_{,L}
=0.
\label{eq:eom1}
\end{align}
We now make the field redefinition
\begin{align}
\psi = \phi + \epsilon F_{,L}[\phi,(\nabla \phi)^2,\square \phi].
\label{eq:redef}
\end{align}
Rewriting the equation of motion (\ref{eq:eom1}) in terms of $\psi$
yields
\begin{align}
\square \psi - U'(\psi) + \epsilon U''(\psi) F_{,L} + \epsilon
F_{,\phi} - 2 \epsilon \nabla_\alpha ( F_{,K} \nabla^\alpha \psi) =
O(\epsilon^2),
\label{eq:psieqn}
\end{align}
where the arguments of $F_{,\phi}, F_{,L}$ and $F_{,K}$ are now
$[\psi,(\nabla \psi)^2,\square \psi]$.

We now apply the reduction of order procedure to the equation of
motion given by Eqs.\ (\ref{eq:redef}) and (\ref{eq:psieqn}),
 that is, we substitute in the zeroth order equation of motion $\square \psi = U'(\psi)$.
The field redefinition (\ref{eq:redef}) gets replaced by the following field redefinition which does not involve higher derivatives:
\begin{align}
\psi = \phi + \epsilon F_{,L}[\phi,(\nabla \phi)^2,U'(\phi)] + O(\epsilon^2).
\label{eq:redef1}
\end{align}
The equation of motion (\ref{eq:psieqn}) is unchanged, except that the arguments of $F_{,\phi}, F_{,L}$ and $F_{,K}$ are now
$[\psi,(\nabla \psi)^2,U'(\psi)]$.  This equation of  motion can be obtained from the action
\begin{align}
  S = \int d^4 x \sqrt{-g} \left\{ - \frac{1}{2} (\nabla \psi)^2 -
    U(\psi) + \epsilon F[\psi, (\nabla \psi)^2, U'(\psi) ]  \right\}.
\label{eq:reduced}
\end{align}
Finally we rewrite this action in terms of $\phi$ using the change of variable (\ref{eq:redef1}).  The result is
\begin{align}
  S = \int d^4 x \sqrt{-g} \left\{ - \frac{1}{2} (\nabla \phi)^2 -
    U(\phi) + \epsilon F[\phi, (\nabla \phi)^2, U'(\phi) ]
+ \epsilon [ \square \phi - U'(\phi) ] F_{,L}[\phi,(\nabla \phi)^2, U'(\phi)] \right\}.
\label{eq:reduced1}
\end{align}
Note that although this action contains second order derivatives,
the corresponding equations of motion contain derivatives only up to
second order, that is, the theory is no longer a ``higher derivative''
theory \cite{Deffayet:2010qz}.  The final, reduced-order action
(\ref{eq:reduced1}) is of
the form (\ref{eq:hatF}) claimed above.

The final result (\ref{eq:reduced1}) shows explicitly that the method
of reducing order directly in the action used in Refs.\
\cite{Weinberg2008,Watson2010} is not correct.  Applying this
procedure to the action (\ref{eq:hd}) would yield the first three
terms in the action (\ref{eq:reduced1}), but not the fourth term.

\subsection{Method of Integrating Out the Additional Fields}

We next show that the same result (\ref{eq:reduced1}) can be obtained
by integrating out the new degrees of freedom that are associated with
the higher derivative terms.  Starting from the action (\ref{eq:hd}),
we introduce an auxiliary scalar field $\psi$ and consider the action
\begin{align}
  S[\phi,\psi] = \int d^4 x \sqrt{-g} \left\{ - \frac{1}{2} (\nabla \phi)^2 -    U(\phi) + \epsilon F[\phi, (\nabla \phi)^2, \psi ]
+ \epsilon (\square \phi - \psi) F_{,L}[\phi,(\nabla \psi)^2,\psi].
 \right\},
\label{eq:hd1}
\end{align}
The equation of motion for $\psi$ from this action is $\psi = \square
\phi$, assuming $F_{,LL} \ne 0$, and substituting this back into the
action (\ref{eq:hd1}) yields the action (\ref{eq:hd}).  Thus the two
actions are equivalent classically.

We now proceed to integrate out the field $\psi$, at tree level, i.e.,
classically.  The equation of motion for $\phi$ is $\psi = U'(\phi) +
O(\epsilon)$, and substituting this back the action (\ref{eq:hd1})
gives the same result (\ref{eq:reduced1}) as was obtained from the
reduction of
order method.

\subsection{Field Redefinition Method}

We next turn to a discussion of the method we use to eliminate higher
derivative terms in the body of the paper, using perturbative field
redefinitions.  That method is not generally applicable, but when it
can be used, it is equivalent to the method of reduction of order
(steps 1-3 above), as we now show.
We start with an action of the form (\ref{eq:hd}), with the function
$F$ chosen to be of the form
\begin{align}
F(\phi,K,L) = g(\phi,K) + [L - U'(\phi)] h(\phi,K,L),
\label{eq:eg}
\end{align}
for some functions $g$ and $h$.
This is the most general form of $F$ for which the field redefinition
method can  be used to eliminate the higher derivatives, and is
sufficiently general to encompass the cases used in the body of the
paper.  First, we apply the reduction of order method.  Inserting the
formula (\ref{eq:eg}) into Eq.\ (\ref{eq:hatF}) shows that the
reduced-order action is characterized by the function ${\hat F}$ given
by
\begin{align}
{\hat F}(\phi,K,L) = g(\phi,K) + [L - U'(\phi)] h[\phi,K,U'(\phi)].
\label{eq:eg1}
\end{align}
However, the same result is obtained by starting with the action given
by Eqs.\ (\ref{eq:hd}) and (\ref{eq:eg}) and performing the field
redefinition
\begin{align}
\phi \to \phi + \epsilon h[\phi,(\nabla \phi)^2,U'(\phi)] -
\epsilon h[\phi,(\nabla \phi)^2,\square \phi].
\end{align}
This shows the reduction of order and field redefinition methods are
equivalent.

We now give a more general and abstract argument for the equivalence,
valid for any field content.  Suppose we have a theory containing
higher derivative terms in the action, proportional to $\epsilon$.
Suppose that we can find a linearized field redefinition, involving
higher derivatives, that has the effect of eliminating all higher
derivative terms from the action.  We can then consider this process
in reverse: starting from a theory which is not higher derivative,
by making a linearized field redefinition we obtain another theory
which has higher derivative terms, proportional to $\epsilon$.
However, the change in the action induced by the field
redefinition must be proportional to the equations of motion.
Hence, these higher derivative terms will be eliminated by applying
Weinberg's method of substituting the zeroth order equations of motion
into the $O(\epsilon)$ terms in the action.
As we have discussed, Weinberg's procedure is valid up to a field
redefinition of the type (\ref{eq:redef1})
which does not change the differential order.

\ifx\doejects\undefined
\else
        \eject
\fi
\section{Comparison with Previous Work}
\label{app:compare}

In this Appendix we compare our analysis and results to those of Park,
Watson and Zurek \cite{Watson2010}, who perform a similar computation
with similar motivation, but obtain a somewhat different final result
[Eq.\ (1) of their paper].  The main differences that arise are:

\begin{itemize}

\item They work throughout in the Jordan frame, whereas we work in the
  Einstein frame.  This is a minor difference which only affects the
  appearance of the computations and results, since it is always
  possible to translate from one frame to another.

\item As discussed in the Introduction and in Appendix
  \ref{app:backsubs}, they use Weinberg's method
  of eliminating the higher derivative terms, consisting of
  substituting the zeroth order equations of motion into the higher
  derivative terms in the action, whereas we use the field
  redefinition method.  The two methods are not equivalent for a given
  specific theory
  with specific coefficients, but are equivalent for the purpose of
  determining a general class of theories.

\item After eliminating higher derivative terms, their result is
an action [Eq.\ (5) of their paper] that contains eleven functions of
the scalar field, whereas our corresponding result
(\ref{eq:details:finalactionJ}) has only
nine free functions.  However, this is a minor difference: their
function $Z(\phi)$ can be eliminated by redefining the scalar field to
attain canonical normalization, and their function $f(\phi)$ can be
eliminated by the transformation used in step 7 in Sec.\
\ref{sec:derivation1} above.

\item Another minor difference is that in their analysis they
have in their action a Weyl squared term $\propto C_{\alpha\beta\gamma\delta}
 C^{\alpha\beta\gamma\delta}$, which is unaffected by any of the
 transformation they make to the action.  This Weyl squared term
gives rise to higher derivative terms in the equation of motion that
are associated with ghost-like additional degrees of
freedom \cite{Chiba2005}.
In our analysis the Weyl squared term is
replaced by the Gauss-Bonnet term, which is not a higher derivative
term, because it would be a topological term if it were not for the
$\phi$-dependent prefactor.

\item Aside from the above minor differences, our result
  (\ref{eq:details:finalaction}) is equivalent to the result given in
  Eq.\ (5) of their paper.
Two major
 differences arise subsequently in the estimates of the
scalings for the
coefficients of the operators in the Lagrangian.

First, Park {\it et al.} use the standard effective theory scaling
rule wherein an operator of dimension $4+n$ has a coefficient $\sim
\Lambda^{-n}$, where $\Lambda$ is the cutoff.  As discussed in Sec.\
\ref{sec:scaling} above, this corresponds to placing no restrictions
on the theory that applies above the cutoff scale $\Lambda$.  By
contrast, our approach does place restrictions on the physics at
scales above $\Lambda$, and yields the modified scaling rule
(\ref{rule}).  As a consequence, our cutoff $\Lambda$ (which we denote by $M$
in the body of the paper) can be taken all
the way down to the Hubble scale $H_0 \sim 10^{-33} \,$ eV, whereas
their cutoff must be larger than $\sim \sqrt{H_0 \mpl} \sim 10^{-3} \,
$ eV.

Second, Park {\it et al.}\ actually assume separate cutoffs for the
gravitational, matter and scalar sectors of the theory, and estimate
how each of their coefficients scale as functions of these three
cutoffs.
We do not understand completely their method of
derivation of these scalings, but we do note that some of their scaling
estimates are
inconsistent with how the coefficients transform into one another
under field redefinitions as discussed in Sec.\ \ref{sec:techs} above.
They then proceed to drop some terms which their scalings indicate are
subdominant, and arrive at a final action [Eq.\ (1) in their paper]
which differs from ours, being parameterized by three free functions
rather than nine.

\end{itemize}

\ifx\doejects\undefined
\else
        \eject
\fi
\section{Equations of Motion for Reduced Theory}
\label{app:eoms}

In this Appendix we compute the equations of motion for our final
action (\ref{eq:details:finalaction}), with the $e_1$ and $e_2$ terms
omitted.  We start by using a transformation of the form
(\ref{eq:tr00a}) with ${\tilde \beta}_2 = -2 e^{-2 \alpha} b_2$. This
yields the action
\begin{align}
S = \int d^4 x \sqrt{-g}  &\left\{ \frac{m_p^2}{2} R - \frac{1}{2} (\nabla \phi)^2 - U(\phi) + a_1 (\nabla \phi)^4 + c_1 G^{\mu \nu} \nabla_\mu \phi \nabla_\nu \phi \right.
\nonumber \\
& \left. + d_3 \left( R^2 - 4 R^{\mu \nu} R_{\mu \nu} + R_{\mu \nu \sigma \rho} R^{\mu \nu \sigma \rho} \right) + d_4 \epsilon^{\mu \nu \lambda \rho} \tensor{C}{_{\mu \nu}^{\alpha \beta}} C_{\lambda \rho \alpha \beta}
\right\} \nonumber \\
&+ S_{\rm m} \left[e^{\alpha(\phi)} g_{\mu \nu} \left( 1 + \beta
    (\nabla \phi)^2 \right), \psi_{\rm m} \right]. \label{action}
\end{align}
Here we have defined $\beta = 2 e^{-2 \alpha} b_2$; this was denoted
$\beta_2$ in the body of the paper. We have also set $\epsilon=1$ for
simplicity.
The representation (\ref{action}) is more convenient than
(\ref{eq:details:finalaction}) for computing the equations of motion
since it avoids varying of the stress-energy tensor.

Next, we vary the matter action in Eq.\ (\ref{action}) using the
definition (\ref{eq:Tabdef}) of the stress energy tensor $T_{\mu\nu}$
and the definition (\ref{eq:JordanMetric}) of the Jordan metric ${\bar g}_{\mu\nu}$.
This yields
\begin{align}
  \delta S_m = - \frac{1}{2} \int d^4 x \sqrt{-g} e^{2 \alpha} & \left\{ \delta g^{\mu \nu} \left[ T_{\mu \nu} + 2 T_{\mu \nu} \beta (\nabla \phi)^2 - \beta T \nabla_\mu \phi \nabla_\nu \phi \right] \right.
\nonumber \\
  & \left.
  + \delta \phi \left[
  - \alpha^\prime T
  + 2 \alpha^\prime \beta T (\nabla \phi)^2
  + \beta^\prime T (\nabla \phi)^2
  + 2 \beta \nabla_\mu T \nabla^\mu \phi
  + 2 \beta T \square \phi
  \right] \right\}.
\end{align}
Combining this with the variation of the gravitational action gives
the equations of motion
\begin{align}
  \square \phi = {}& U^\prime (\phi) - \frac{1}{2} e^{2 \alpha} \alpha^\prime T + 4 a_1 \left[(\nabla \phi)^2 \square \phi + 2 \nabla_\mu \nabla_\nu \phi \nabla^\mu \phi \nabla^\nu \phi \right] + 3 a_1^\prime (\nabla \phi)^4 + c_1^\prime G^{\mu \nu} \nabla_\mu \phi \nabla_\nu \phi
\nonumber \\
  {}& + 2 c_1 G^{\mu \nu} \nabla_\mu \nabla_\nu \phi - d_3^{\prime} \left( R^2 - 4 R^{\mu \nu} R_{\mu \nu} + R_{\mu \nu \sigma \rho} R^{\mu \nu \sigma \rho} \right) - d_4^\prime \epsilon^{\mu \nu \lambda \rho} \tensor{C}{_{\mu \nu}^{\alpha \beta}} C_{\lambda \rho \alpha \beta}
\nonumber \\
  {}& + \frac{1}{2} e^{2 \alpha} \left[
    2 \alpha^\prime \beta T (\nabla \phi)^2
  + \beta^\prime T (\nabla \phi)^2
  + 2 \beta \nabla_\mu T \nabla^\mu \phi
  + 2 \beta T \square \phi
  \right],
\end{align}
and
\begin{align}
  m_p^2 G_{\mu \nu} = {}& e^{2 \alpha} T_{\mu \nu} + \nabla_\mu \phi \nabla_\nu \phi - \left[\frac{1}{2} (\nabla \phi)^2 + U(\phi) \right] g_{\mu \nu} - 4 a_1 (\nabla \phi)^2 \nabla_\mu \phi \nabla_\nu \phi + a_1 (\nabla \phi)^4 g_{\mu \nu}
\nonumber \\
  & + g_{\mu \nu} c_1 G^{\sigma \lambda} \nabla_\sigma \phi \nabla_\lambda \phi - 4 c_1 R_{\sigma (\mu} \nabla_{\nu )} \phi \nabla^\sigma \phi + c_1 (R_{\mu \nu} (\nabla \phi)^2 + R \nabla_\mu \phi \nabla_\nu \phi)
\nonumber \\
  & - g_{\mu \nu} \nabla_\sigma \nabla_\lambda (c_1 \nabla^\sigma \phi \nabla^\lambda \phi) + g_{\mu \nu} \square [c_1 (\nabla \phi)^2] + 2 \nabla_\lambda \nabla_{(\mu} (c_1 \nabla_{\nu)} \phi \nabla^\lambda \phi ) - \nabla_\mu \nabla_\nu [c_1 (\nabla \phi)^2]
\nonumber \\
  & - \square(c_1 \nabla_\mu \phi \nabla_\nu \phi) + 2 R \nabla_\mu \nabla_\nu d_3 - 2 g_{\mu \nu} R \square d_3 + 4 R_{\mu \nu} \square d_3 - 8 \tensor{R}{^\sigma_{(\mu}} \nabla_{\nu)} \nabla_\sigma d_3
\nonumber \\
  & + 4 g_{\mu \nu} R_{\sigma \rho} \nabla^\sigma \nabla^\rho d_3 + 4 R_{\rho \mu \nu \sigma} \nabla^\rho \nabla^\sigma d_3 + 16 C_{\mu \nu} + 2 e^{2 \alpha} T_{\mu \nu} \beta (\nabla \phi)^2 - e^{2 \alpha} \beta T \nabla_\mu \phi \nabla_\nu \phi. \label{eq:tensoreom}
\end{align}
Here the tensor $C_{\mu \nu}$ comes from the Chern-Simons term, and is
defined by
\begin{align}
  C^{\mu \nu} = (\nabla_\sigma d_4) \epsilon^{\sigma \lambda \rho(\mu} \nabla_\rho \tensor{R}{^{\nu)}_\lambda} + (\nabla_\sigma \nabla_\lambda d_4) \tensor[^\star]{R}{^{\lambda(\mu \nu)\sigma}} \label{eq:chernsimons}
\end{align}
where $\tensor[^\star]{R}{^{\mu \nu \sigma \lambda}} =
\epsilon^{\sigma \lambda \rho \tau} \tensor{R}{^{\mu \nu}_{\rho \tau}}
/ 2$. Note that the zeroth order terms involving the stress-energy
tensor depend implicitly on $\beta$ through the expression for the
Jordan metric given in Eq.\ (\ref{action}).

The terms involving $c_1$ are written in the most compact manner we could find. Although it looks unlikely, the higher order derivatives in these terms do cancel; the full expansion of these terms is
\begin{align}
  & 2 c_1 g_{\mu \nu} R^{\sigma \lambda} \nabla_\sigma \phi \nabla_\lambda \phi - \frac{1}{2} c_1 g_{\mu \nu} R (\nabla \phi)^2 - 4 c_1 R_{\sigma (\mu} \nabla_{\nu)} \phi \nabla^\sigma \phi + c_1 R_{\mu \nu} (\nabla \phi)^2 + c_1 R \nabla_\mu \phi \nabla_\nu \phi
\nonumber \\
  &+ g_{\mu \nu} \left[ c_1^\prime \nabla_\sigma \phi \nabla_\lambda \phi \nabla^\sigma \nabla^\lambda \phi + c_1 \nabla_\sigma \nabla_\lambda \phi \nabla^\sigma \nabla^\lambda \phi - c_1^\prime (\nabla \phi)^2 \square \phi - c_1 (\square \phi)^2 \right]
\nonumber \\
  & - 2 c_1 \nabla_\sigma \nabla_\mu \phi \nabla^\sigma \nabla_\nu \phi - 2 c_1^\prime \nabla_\sigma \phi \nabla_{(\mu} \phi \nabla_{\nu)} \nabla^\sigma \phi + c_1^\prime \nabla_\mu \nabla_\nu \phi (\nabla \phi)^2 + c_1^\prime \nabla_\mu \phi \nabla_\nu \phi \square \phi
\nonumber \\
  & + 2 c_1 \nabla_\mu \nabla_\nu \phi \square \phi + 2 c_1 \nabla^\lambda \phi \nabla^\sigma \phi R_{\sigma \mu \nu \lambda}.
\end{align}

\ifx\doejects\undefined
\else
        \eject
\fi
\section{Scaling of Coefficients Obtained by Integrating Out Pseudo-Nambu-Goldstone Fields}
\label{app:integrateout}

In this Appendix we give some more details of the derivation
discussed in Sec.\ \ref{sec:scaling} of the scaling
of the coefficients of the operators in the Lagrangian.
We divide the PNGB fields $\Phi^A$ into two groups, a set $\chi^a$
with mass $\sim H_0$ and a set $\psi^\Gamma$ with mass $\sim M$, where
$M \gg H_0$:
\begin{align}
\Phi^A = (\chi^a,\psi^\Gamma).
\end{align}
We assume an action for these fields of the form
\begin{align}
S = \int d^4 x \sqrt{-g} \left\{ \frac{1}{2} R - \frac{1}{2}
q_{AB}(\Phi^A) \nabla_\mu \Phi^A \nabla_\nu \Phi^B g^{\mu\nu}
- H_0^2 V\left( \chi^a, \frac{M}{H_0} \psi^\Gamma \right) \right\}.
\label{action6}
\end{align}
This is the same as the action
(\ref{eq:S0M2}) of Sec.\ \ref{sec:scaling} above, except that an extra
factor has been inserted into the potential to make the $\psi^\Gamma$
fields have mass $\sim M$ rather than $\sim H_0$, and we have
specialized to units where $\mpl=1$.
We assume that the target space coordinates have been chosen so that the
potential is minimized at $\psi^\Gamma = 0$, i.e.
\begin{align}
V_{,\Gamma} = 0
\label{eq:surface}
\end{align}
at $\psi^\Gamma=0$.

We now want to let $M$ become large and integrate out the fields
$\psi^\Gamma$ at tree level.
This can be done by using Feynman diagrams and using power
counting\footnote{We note that Burgess
  \textit{et al}. \protect{\cite{Burgess:2009ea}} write down a scaling rule
in their Eqs.\ (2.3) and (2.5)
which is identical to
our scaling rule (\ref{rule}) except that it is suppressed by an
overall factor of
$M^2/\mpl^2$ for $d > 2$, where $d$ is the number of derivatives.
They say in their footnote 2 that this rule comes from integrating
out a PNGB field of mass $M$. However we find that the detailed power
counting calculations given in the second example in their Sec.\ 2.2
actually
yield our scaling rule rather than theirs.},
as in Ref.\ \cite{Burgess:2009ea}.  Alternatively and more simply,
it can be done by writing out the equations of motion for the fields
$\psi^\Gamma$ and invoking an adiabatic approximation.
At zeroth order in $1/M$, the theory obtained for the fields
$\chi^a$ is a nonlinear sigma model where the potential is just the
potential of the action (\ref{action6}) evaluated on the surface
$\psi^\Gamma=0$, and the target space metric is just the metric
induced on the surface from the metric $q_{AB}$.

To obtain the higher order corrections we can proceed as follows.
The equation of motion for the fields $\psi^\Gamma$ is
\begin{align}
\square \psi^\Sigma + \Gamma^\Sigma_{ab} {\vec \nabla} \chi^a \cdot
{\vec \nabla} \chi^b +
\Gamma^\Sigma_{\Theta\Upsilon} {\vec \nabla} \psi^\Theta \cdot {\vec
  \nabla} \psi^\Upsilon
+ 2 \Gamma^\Sigma_{a \Theta} {\vec \nabla} \chi^a \cdot {\vec \nabla}
\psi^\Theta = H_0^2 q^{\Sigma a} V_{,a} + H_0 M q^{\Sigma\Theta} V_{,\Theta}.
\end{align}
Here the connection coefficients are those of the target space metric
$q_{AB}$.  We next expand this equation to linear order in
$\psi^\Gamma$ and use the condition (\ref{eq:surface}) to obtain
\begin{align}
&&\square \psi^\Sigma + \left[ \Gamma^\Sigma_{ab,\Theta} {\vec \nabla} \chi^a \cdot
{\vec \nabla} \chi^b
- H_0^2 q^{\Sigma a}_{\ \ \ ,\Theta} V_{,a} - M^2 q^{\Sigma\Upsilon} V_{,\Upsilon\Theta}
\right] \psi^\Theta
\nonumber \\
&&
+ 2 \Gamma^\Sigma_{a \Theta} {\vec \nabla} \chi^a \cdot {\vec \nabla}
\psi^\Theta = - \Gamma^\Sigma_{ab} {\vec \nabla} \chi^a \cdot
{\vec \nabla} \chi^b +H_0^2 q^{\Sigma a} V_{,a},
\end{align}
where all the metric coefficients, connection coefficients and
their derivatives are evaluated at $\psi^\Gamma=0$.
Now in the large $M$ or adiabatic limit, the dominant term on the left
hand side will be the term proportional to $M^2$, and dropping the
other terms gives a simple algebraic equation for the leading order
contribution to $\psi^\Gamma$:
\begin{align}
\left[ q^{\Sigma\Upsilon} V_{,\Upsilon\Theta} \right] \psi^\Theta
= \frac{1}{M^2} \left[ \Gamma^\Sigma_{ab} {\vec \nabla} \chi^a \cdot
{\vec \nabla} \chi^b -H_0^2 q^{\Sigma a} V_{,a} \right].
\label{eq:soln1}
\end{align}
Substituting the solution given by Eq.\ (\ref{eq:soln1}) into the
action (\ref{action6}) gives the required, $O(1/M^2)$ corrections to
the action.  The first term on the right hand side of Eq.\ (\ref{eq:soln1})
will give nonlinear corrections to the kinetic energy.
(We assume that the second fundamental form or extrinsic curvature of
the surface $\psi^\Gamma=0$ is nonzero, otherwise these
corrections would vanish.)

As a simple example, consider the theory
\begin{align}
{\cal L} = - \frac{1}{2} (\nabla \chi)^2 - \frac{1}{2} (\nabla \psi)^2
- \frac{1}{2} M^2 \psi^2 + \psi (\nabla \chi)^2/\mpl.
\end{align}
The equation of motion for $\psi$ is $\square \psi - M^2 \psi =
(\nabla \chi)^2/\mpl$ with leading order solution $\psi = -(\nabla
\chi)^2 / (\mpl M^2)$.  The corresponding corrections to the action
for $\chi$ scale as $(\nabla \chi)^4/(\mpl^2 M^2)$, in agreement with
Eq.\ (\ref{rule}).  The scaling (\ref{rule}) of other operators can be
derived similarly.

\end{document}